\documentclass[
reprint,
superscriptaddress,
amsmath,
amssymb,
aps,
prapplied,
floatfix
]{revtex4-2}

\usepackage[T1]{fontenc}
\usepackage{verbatim}
\usepackage{xcolor}
\usepackage{graphicx}
\usepackage{dcolumn}
\usepackage{amsthm}
\usepackage{bm}
\usepackage{tikz}
\usetikzlibrary{arrows.meta,positioning,shapes.geometric}
\usepackage{hyperref}

\newcommand{\twopanelgraphic}[1]{%
  \begin{tikzpicture}
    \node[inner sep=0pt,outer sep=0pt,anchor=south west] (panels) at (0,0)
      {\includegraphics[width=0.85\textwidth]{#1}};
    \begin{scope}[x={(panels.south east)},y={(panels.north west)}]
      \node[anchor=north,font=\bfseries] at (0.27,-0.02) {(a)};
      \node[anchor=north,font=\bfseries] at (0.77,-0.02) {(b)};
    \end{scope}
  \end{tikzpicture}%
}

\begin{document}

	\title{MUTACO: Simulator-Efficient Response Matching in Stochastic Nonlinear Systems via Compatibility-Aware Sequential Design}

	\author{Ege Karadeniz}
	\email{ege.karadeniz@ug.bilkent.edu.tr}
	\affiliation{Department of Physics, Bilkent University, Ankara 06800, Türkiye}

	\author{\.{I}lter Onat Korkmaz}
	\affiliation{Department of Electrical and Computer Engineering, Texas A\&M University, College Station, Texas 77843, USA}
	\affiliation{Department of Electrical and Electronics Engineering, Bilkent University, Ankara 06800, Türkiye}

	\author{Umut Can Turhan}
	\affiliation{Department of Physics, Bilkent University, Ankara 06800, Türkiye}

	\author{Cem Tekin}
	\affiliation{Department of Electrical and Electronics Engineering, Bilkent University, Ankara 06800, Türkiye}

	\author{Aydın Cem Keser}
	\email{aydin.keser@bilkent.edu.tr}
	\affiliation{Department of Physics, Bilkent University, Ankara 06800, Türkiye}

	\date{\today}

	\begin{abstract}
		Many physical inverse problems are mediated by expensive stochastic simulators
		for which response gradients, adjoint sensitivities, or a differentiable
		simulator interface are unavailable.
		We introduce \emph{Multiple Target-based Confident Acquisition} (MUTACO), an
		uncertainty-aware sequential-design method for response matching that requires
		only noisy forward evaluations and learns the multicomponent response
		together with its predictive uncertainty.
		We develop and evaluate MUTACO using a heterogeneous network of stochastic phase oscillators. The inverse task is to infer statistical properties of the local
		potential and interactions from macroscopic rotation and synchronization curves
		under multiple driving conditions.
		Rather than requiring unique recovery of these statistical properties, MUTACO
		searches for parameter configurations whose predicted rotation and synchronization
		responses are compatible with the target response while accounting for
		predictive uncertainty.
		This response-centered formulation is appropriate when experimentally accessible
		observables characterize ensemble statistics rather than a particular microscopic
		realization.
		Under strictly limited simulation budgets, MUTACO achieves the lowest held-out
		response error among the tested adaptive-search, space-filling, and
		simulation-based inference methods, while also yielding lower parameter-recovery
		error than the compared methods.
		These results demonstrate that response-targeted sequential design is a
		simulator-efficient approach to inverse modeling from finite noisy forward
		evaluations and suggest broader applicability to stochastic physical systems with correlated 
		multicomponent responses and weakly constrained parameter directions.
	\end{abstract}

	\maketitle

	%\tableofcontents

	\section{Introduction}
	Inverse problems in nonlinear dynamical systems play a central role in
	contemporary physics and scientific computing~\cite{tarantola_inverse,stuart_bayesian_inverse}. In many settings, the forward dynamics can be modeled explicitly, while recovering the underlying parameters from macroscopic observations remains a challenging inverse problem~\cite{arridge_inverse_review}.
	This difficulty is amplified in stochastic and disordered systems, where
	stochastic fluctuations, quenched heterogeneity, or unresolved microscopic
	degrees of freedom make observable responses variable even at fixed
	parameter values. Such scenarios arise in climate and atmospheric
	modeling~\cite{wunsch_ocean_inverse}, data-driven discovery of nonlinear
	dynamics~\cite{brunton_sindy}, materials
	design~\cite{lookman_active_learning_materials}, quantum-system
	characterization and Hamiltonian learning~\cite{gebhart_learning_quantum},
	and networks of coupled oscillators~\cite{acebron_kuramoto_review}.
	In these systems, inversion must contend not only with nonlinearity but also
	with randomness that obscures the relation between parameters and observables.

In practice, many such inverse problems are mediated by expensive numerical
simulators rather than analytically tractable forward models.
Gravitational-wave parameter estimation provides one example, where inference
for compact-binary signals can require repeated evaluations of computationally
expensive waveform models~\cite{bhardwaj_gw_ssbi}.\par

		Related challenges arise in heterogeneous networks of biological and
		artificial neural oscillators, where variability in local dynamics,
		interaction strengths, and stochastic forcing affects collective
		activity and synchronization
		~\cite{winfree_biological_time,tanaka_physical_reservoir,
			romera_coupled_sto,kanao_sto_reservoir}.
		A complementary approach reconstructs effective statistical
		interactions from neural population activity using maximum-entropy
		(inverse-Ising) models~\cite{schneidman_bialek_neural_ising,
			meshulam_bialek_2025}.
		Such approaches illustrate how collective measurements can constrain
		reduced models without resolving detailed cellular mechanisms.
		For stochastic dynamical networks, inverse characterization can
		likewise focus on macroscopic responses rather than the detailed
		reconstruction of individual units.
		Repeated forward evaluations at the same parameter vector may then
		produce different responses, while the corresponding likelihood
		can be unavailable or prohibitively expensive to evaluate.\par\par

	Fully Bayesian treatments based on Markov chain Monte Carlo
	~\cite{neal_mcmc_review,robert_casella_mcmc} or exchange Monte Carlo
	~\cite{hukushima_exchange_mc} can require large numbers of forward evaluations
	when likelihood evaluation itself depends on an expensive simulator. More recently, simulation-based inference (SBI) methods have sought to
	reduce this cost using neural density estimators~\cite{papamakarios_murray_npe,papamakarios_snl,cranmer_sbi_review,
		lueckmann_snpe,BoeltsDeistler_sbi_2025, tejero-cantero2020sbi}.
	However, accurately characterizing a posterior distribution is not always
	synonymous with accurately reproducing a desired macroscopic response.
	This distinction is particularly relevant when the quantities of interest
	are collective observables generated by a stochastic many-body system.

In this work, we focus on the inverse characterization of disordered
networks of stochastic nonlinear phase oscillators.
These networks provide an idealized phenomenological model of
interacting neural-like units, combining local
nonlinear restoring forces, heterogeneous phase coupling, external
driving, and stochastic fluctuations
\cite{acebron_kuramoto_review,strogatz_sync_review}.
The task is to infer statistics of the local potential and interactions
from macroscopic rotation and synchronization measurements under
multiple driving conditions.
Fig.~\ref{fig:inverse_problem_schematic} summarizes this
simulator-mediated inverse experiment.

	Within this driven setting, the macroscopic oscillator network response ranges from
	incoherent fluctuations to partially synchronized and rotating states, while
	disorder and noise induce substantial variability in the measured observables.
	Rather than inferring every microscopic parameter individually, we describe
	the system through a small set of ensemble statistics  that govern the
	underlying potential and interactions.
	The inverse problem is then to identify ensemble-parameter configurations
	whose macroscopic rotation and synchronization responses reproduce prescribed
	target behavior across multiple driving conditions. This formulation reflects
	a broader class of stochastic systems whose large-scale behavior can be
	characterized by a limited number of ensemble parameters while remaining
	computationally expensive to simulate.

We formulate inversion as the direct search for ensemble-parameter
configurations that reproduce a prescribed macroscopic response. In the
oscillator problem above, this means searching over the parameter statistics
$\boldsymbol{\theta}$ for configurations whose disorder-averaged rotation and
synchronization measurements match the target values across the prescribed
driving conditions. Rather than first attempting to characterize a posterior
distribution over the ensemble parameters, we seek parameter values whose
predicted response vector reproduces these target observables.

This viewpoint connects naturally to classical and Bayesian experimental
design~\cite{fedorov_experimental_design,chaloner_bayesian_design}, as well as
to Bayesian optimization (BO) and bandit
methods~\cite{auer_bandits,brochu_bo_review,srinivas_gpucb,demirel2022escada},
where uncertainty-aware acquisition strategies guide the sampling of expensive
black-box functions, including simulation-driven physical
design~\cite{elkhettabi_bo_cfd,shen_device_bo}. Related work extends this
perspective to function-valued simulator outputs using neural-operator
surrogates~\cite{oliveira_function_space_ts}. Response-matching problems can
either reduce the discrepancy from the target to a scalar objective or retain
the target-vector structure explicitly, as in target-vector and
composite-function Bayesian
optimization~\cite{uhrenholt2019target,plock_target_vector_2022,
	astudillo_composite_bo}. The latter provides the natural starting point for
the response-focused inverse problem considered here.

When richer information about the forward model is available, physical
optimization can additionally exploit response sensitivities. Gradients may,
for example, be extracted directly from controlled experimental perturbations
~\cite{boon_gradient_in_materia}, while differentiable physical models permit
efficient adjoint inverse design. Recent nanophotonic approaches illustrate
the latter regime by optimizing structured optical transformations and
realizing them through full-wave electromagnetic inverse design with adjoint
sensitivities~\cite{tegin_scattering_inverse,tegin_deep_inverse}.

To address the complementary regime, in which response sensitivities and
model-level access are unavailable or not exposed to the optimizer, we
introduce Multiple Target-based Confident Acquisition (MUTACO), a
sequential-design method for response-focused inversion. MUTACO requires only
stochastic input--output evaluations of the forward simulator; response
gradients, adjoint equations, governing equations, and simulator internals are
not required. Within the target-vector perspective described above, MUTACO is
distinguished by using the structure of the multicomponent response and its
predictive uncertainty directly during sequential acquisition.

MUTACO builds a correlated multitask Gaussian-process surrogate for the full
vector of rotation and synchronization responses, allowing correlations among
observables and driving conditions to be learned together with their posterior
uncertainty. For each candidate $\boldsymbol{\theta}$, the posterior defines an
uncertainty-dependent compatibility criterion relative to the prescribed
target. Target-directed refinement is restricted to candidates satisfying this
criterion, with a logarithmic barrier discouraging approach to its boundary.
When no compatible candidate is identified, the learned posterior covariance
instead drives exploration toward regions of large response uncertainty.

After the simulation budget is exhausted, MUTACO further exploits the
completed surrogate by continuously minimizing the posterior-mean physical
response mismatch over the admissible parameter domain, rather than restricting
the final recommendation to previously simulated points. Thus, the learned
response correlations and predictive uncertainty guide both the allocation of
the limited simulation budget and the extraction of the final response-matching
estimate.

\begin{figure}[t]
	\centering
	\resizebox{\columnwidth}{!}{%
		\begin{tikzpicture}[
			>=Latex,
			font=\small,
			box/.style={
				draw=black!70,
				fill=gray!7,
				rounded corners=1.2pt,
				minimum height=0.72cm,
				minimum width=2.15cm,
				align=center,
				inner sep=3.5pt
			},
			output/.style={box, minimum width=2.35cm},
			arrow/.style={->, line width=0.7pt, draw=black!80},
			stage/.style={font=\footnotesize, text=black!65},
			parameter/.style={font=\footnotesize, align=center}
			]
			
			\node[stage] at (0,1.55) {Controlled input};
			\node[stage] at (3.05,1.55)
			{Oscillator network};
			\node[stage] at (6.55,1.55) {Response};
			
			\node[box] (bias) at (0,0)
			{Bias protocol\\$\mathcal B=\{(F_v,\Phi_u)\}_{u,v}$};
			
			\node[box] (target) at (3.40,0.75)
			{Reference oscillator\\ensemble};
			
			\node[box] (sim) at (3.40,-0.75)
			{Stochastic phase-\\oscillator simulator};
			
			\node[output] (targetout) at (6.55,0.75)
			{Target response\\$\mathbf y^\star$};
			
			\node[output] (simout) at (6.55,-0.75)
			{Simulated response\\$\mathbf y(\boldsymbol{\theta})$};
			
			\coordinate (split) at (1.75,0);
			
			\draw[arrow] (bias.east) -- (split);
			\draw[arrow] (split) |- (target.west);
			\draw[arrow] (split) |- (sim.west);
			
			\draw[arrow] (target.east) -- (targetout.west);
			\draw[arrow] (sim.east) -- (simout.west);
			
			\node[parameter] (theta) at (3.40,-1.90)
			{$\boldsymbol{\theta}=
				(\mu_W,\mu_K,\mu_A,\mathrm{CV}_K)$};
			
			\draw[arrow] (theta.north) -- (sim.south);
			
			\node[font=\small] at (3.05,-2.70)
			{Find $\hat{\boldsymbol{\theta}}\in\Theta$ such that
				$\mathbf y(\hat{\boldsymbol{\theta}})
				\approx\mathbf y^\star$.};
		\end{tikzpicture}%
	}
	\caption{
			Simulator-mediated inverse problem for a heterogeneous network of stochastic phase oscillators.
			A common controlled bias protocol is applied to the
			reference ensemble and the forward simulator.
			In the numerical benchmark, the target response
			$\mathbf y^\star$ is generated synthetically from an
			ensemble parameter vector $\boldsymbol{\theta}^\star$,
			which is withheld during inversion.
			The stochastic simulator evaluates the macroscopic
			rotation and synchronization response
			$\mathbf y(\boldsymbol{\theta})$ for candidate ensemble
			parameters.
			The objective is to reproduce the target response,
			without necessarily recovering the generating
			parameters.}
	\label{fig:inverse_problem_schematic}
\end{figure}

	We evaluate MUTACO on stochastic oscillator-network response matching under strict budgets of only 40 and 100 forward simulations, comparing it with objective-aligned Bayesian optimization, space-filling search, and several simulation-based inference methods. MUTACO achieves the lowest mean held-out response error among the tested methods.
	Notably, although parameter recovery is not its optimization objective,
	MUTACO also achieves lower mean parameter-recovery error than the compared methods. The comparison further shows that proximity in the space of parameter statistics need not imply proximity in the measured collective response, highlighting the distinction between recovering the generating ensemble parameters and reproducing the macroscopic behavior they produce.

	The remainder of this paper is organized as follows.
	Section~\ref{sec:oscillator} introduces the stochastic oscillator-network model,
	its disorder parameterization, and the multibias macroscopic response used in
	the forward problem.
	Section~\ref{sec:inverse} formulates the corresponding inverse problem in
	terms of response reproduction under a finite simulator budget.
	Section~\ref{sec:mutaco} presents the MUTACO compatibility-aware sequential
	design framework.
	Section~\ref{sec:experiments} describes the benchmark targets,
	simulation budgets, comparison methods, randomization protocol, and evaluation
	metrics.
	Section~\ref{sec:results} presents the response-matching, parameter-recovery,
	simulation-based inference, and local-sensitivity results.
	Finally, Section~\ref{sec:discussion} discusses the physical interpretation,
	scope, and limitations of the method, and Sec.~\ref{sec:conclusion} summarizes
	the main conclusions.

	\begin{figure*}[!htp]
		\centering
		\twopanelgraphic{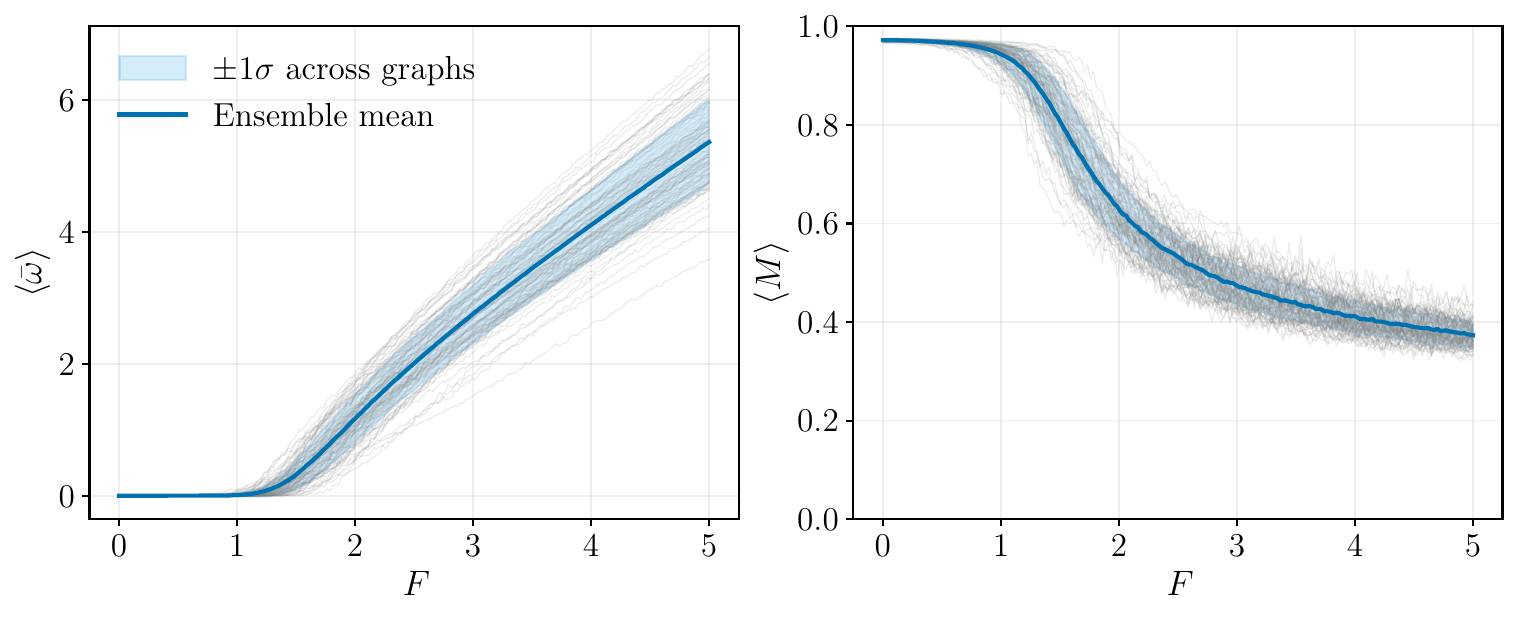}
		\caption{
			Forward response of the coupled network of $N=10$ stochastic
			oscillators for the ensemble-parameter configuration
			$\boldsymbol{\theta}^{\star}
			=(\mu_W,\mu_K,\mu_A,\mathrm{CV}_K)
			=(1.91,1.12,1.16,0.88)$,
			at effective noise intensity $T=0.15$ and phase-bias setting $\Phi=0$.
			Here, $\mu_W$, $\mu_K$, and $\mu_A$ denote the ensemble means of the
			restoring strength, pairwise coupling, and bias scale, respectively,
			while $\mathrm{CV}_K$ denotes the coefficient of variation of the
			coupling strength.
			This response provides the illustrative target reconstructed in
			Fig.~\ref{fig:response_reconstruction}.
			(\textbf{a}) Mean angular velocity
			$\langle\bar{\omega}\rangle$ and
			(\textbf{b}) time-averaged Kuramoto order-parameter magnitude
			$\langle M\rangle$ as functions of the driving force $F$.
			Faint gray traces show 80 disorder realizations, each averaged over
			two independent stochastic repeats.
			The solid blue line denotes the mean across disorder realizations,
			and the light-blue band indicates the mean $\pm$ one standard
			deviation ($\pm1\sigma$). The corresponding $(B=40)$ MUTACO reconstruction is shown in Fig.~\ref{fig:response_reconstruction}.
		}
		\label{fig:forward_response}
	\end{figure*}

	\begin{figure*}[!htp]
		\centering
		\twopanelgraphic{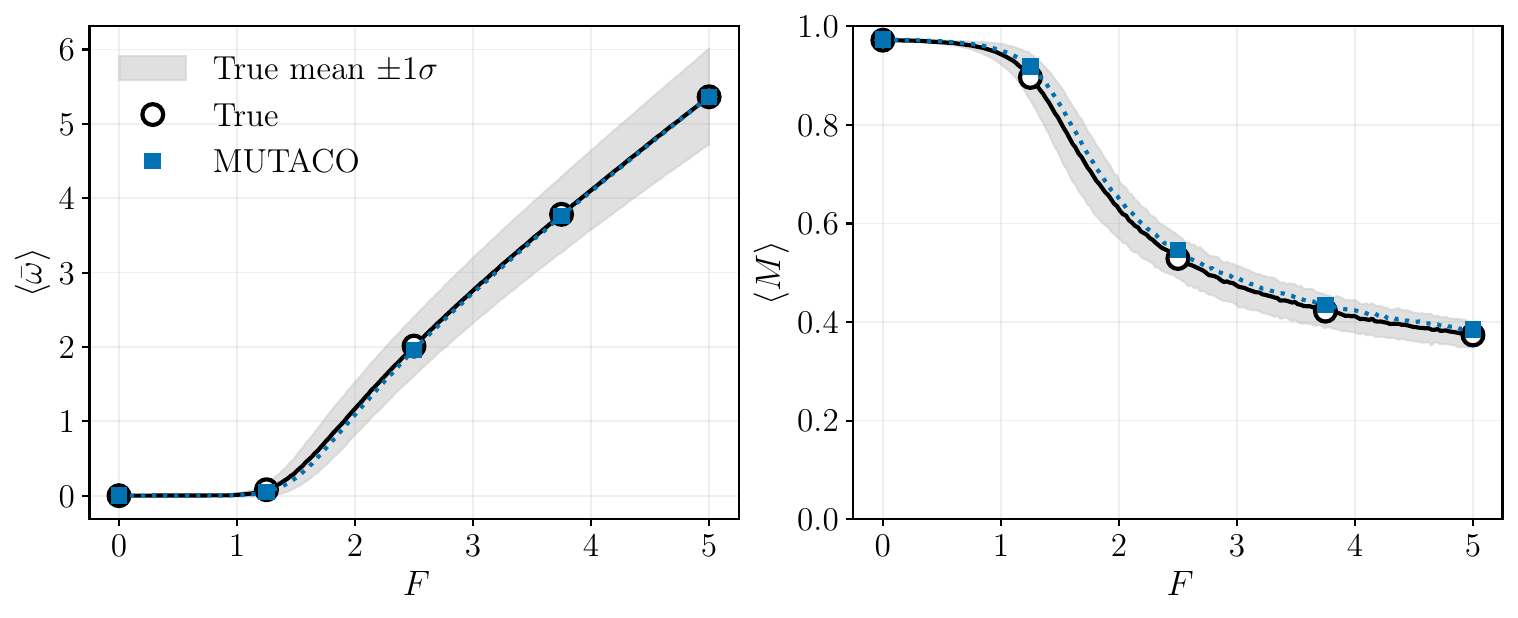}
		\caption{
			MUTACO reconstruction for the illustrative benchmark instance introduced in
			Fig.~\ref{fig:forward_response}, at phase-bias setting $\Phi=0$.
			(\textbf{a}) Mean angular velocity
			$\langle\bar{\omega}\rangle$ and
			(\textbf{b}) time-averaged Kuramoto order-parameter magnitude
			$\langle M\rangle$ as functions of the driving force $F$.
			The black curve reproduces the target response shown in
			Fig.~\ref{fig:forward_response}, while the blue curve is the response
			obtained from the simulator at the $B=40$ MUTACO recommendation
			$\widehat{\boldsymbol{\theta}}
			=(1.95,\,1.18,\,1.13,\,0.41)$,
			with parameter ordering defined in Eq.~\eqref{eq:theta_definition}.
			The gray band denotes the $\pm1\sigma$ variability of the target response
			across 80 disorder realizations, each averaged over two independent
			stochastic repeats.
			Markers indicate the five driving-force values
			$F=0,\ 1.25,\ 2.5,\ 3.75,\ 5$
			used in this $\Phi=0$ slice of the response vector; circles denote the
			target values and squares the MUTACO reconstruction.
			Lines between these sampled conditions visualize the corresponding dense
			forward responses.
			The close agreement illustrates that MUTACO can reproduce the prescribed
			collective response even when the recommended parameters differ appreciably
			from the generating ensemble-parameter vector, most notably in
			$\mathrm{CV}_K$.
			This reconstruction is illustrative; systematic population-level
			response-matching and parameter-recovery performance across the full
			120-target benchmark is quantified in
			Figs.~\ref{fig:primary_response_mae}--\ref{fig:sbi_response_parameter}.
			The local sensitivity analysis below further examines weakly constrained
			parameter directions across a subset of the benchmark.
		}
		\label{fig:response_reconstruction}
	\end{figure*}

\section{Stochastic Oscillator Network Model}
\label{sec:oscillator}

\subsection{Overdamped stochastic dynamics}

We consider a network of $N$ coupled phase oscillators in the
overdamped regime, where inertial effects are neglected.
The phase dynamics are governed by the first-order stochastic
differential equations
\begin{equation}
	\frac{d\phi_i}{d\tau}
	=
	\alpha_i f_i^{(\mathrm{tot})}(\tau)
	+
	\sqrt{2T\alpha_i}\,\eta_i(\tau),
	\label{eq:eom_overdamped}
\end{equation}
where $\phi_i(\tau)$ denotes the phase of oscillator $i$
at time $\tau$, $\alpha_i>0$ is its local mobility,
and $f_i^{(\mathrm{tot})}$ is the total generalized force
acting on it. The mobility $\alpha_i$ is the inverse of
the corresponding viscous damping coefficient.

Here and throughout, energies are measured in units of a
chosen reference scale $E_0$ and time in units of
$(\alpha_0 E_0)^{-1}$, where $\alpha_0$ is a reference
mobility used for nondimensionalization.
Any common dimensional prefactors are absorbed into these
reference scales. Thus, in dimensionless units,
\begin{equation}
	E_0=\alpha_0=k_B=1.
	\label{eq:nondimensional_units}
\end{equation}

The stochastic forcing $\eta_i(\tau)$ is independent
Gaussian white noise satisfying
\begin{equation}
	\langle\eta_i(\tau)\rangle=0,
	\qquad
	\langle\eta_i(\tau)\eta_j(\tau')\rangle
	=
	\delta_{ij}\delta(\tau-\tau').
\end{equation}
The noise amplitude follows the conventional overdamped
Langevin normalization. In a thermal realization, $T$
corresponds to temperature in dimensionless units.
More generally, $T$ serves as an effective stochastic-noise
intensity, without requiring a thermal interpretation
of the fluctuations.

The total deterministic force acting on each oscillator
is decomposed as
\begin{equation}
	f_i^{(\mathrm{tot})}
	=
	f_i^{(\mathrm{loc})}
	+
	f_i^{(\mathrm{coup})}
	+
	f_i^{(\mathrm{bias})}.
\end{equation}

The local restoring force is
\begin{equation}
	f_i^{(\mathrm{loc})}
	=
	-W_i\sin(\phi_i-A_i\Phi),
\end{equation}
where $W_i>0$ sets the local restoring strength and
$\Phi$ is an externally controlled phase-bias parameter.
The coefficient $A_i>0$ sets the local sensitivity to
this bias, so that the preferred phase is shifted
by $A_i\Phi$.

Pairwise interactions between oscillators are described by
\begin{equation}
	f_i^{(\mathrm{coup})}
	=
	-\frac{1}{N}
	\sum_{j=1}^{N}
	K_{ij}\sin(\phi_i-\phi_j).
\end{equation}
The coefficients $K_{ij}=K_{ji}$, with $K_{ii}=0$, are
quenched random variables characterizing the strength
and sign of the interaction between every distinct pair
of oscillators. The resulting network is dense and all-to-all, with heterogeneous, signed Kuramoto-type interactions.
Positive $K_{ij}$ favors pairwise phase alignment,
whereas negative $K_{ij}$ favors an antiphase relationship.
Competing signed interactions can frustrate global phase
alignment, allowing both collective synchronization
and frustrated phase configurations.
The factor $1/N$ provides the usual mean-field normalization, preventing the coupling force from growing with system size when the coupling scale remains fixed.

External driving enters through
\begin{equation}
	f_i^{(\mathrm{bias})}
	=
	A_i F,
\end{equation}
where $F$ denotes a uniform driving force.
We use the same local coefficient $A_i$ for the two
controlled bias channels as a minimal phenomenological
model of correlated bias heterogeneity; this choice
is not intended to represent a universal microscopic
relation between the two couplings in a specific physical
platform. The pair $(F,\Phi)$ therefore defines the
two-dimensional external bias protocol used to probe
the network response.

Phase-reduced descriptions have been used to study
individual neuronal oscillations and collective neural dynamics
without explicitly resolving membrane-voltage waveforms,
action potentials, or microscopic synaptic mechanisms
\cite{stiefel_ermentrout_2016,breakspear_cortical_kuramoto_2010}.
In this phenomenological interpretation, the phase
$\phi_i$ represents progress through an oscillatory cycle,
while sustained phase rotation represents recurrent activity
rather than a directly measured firing rate.
The external drive $F$ represents a sustained effective
input, while the mobility $\alpha_i$ controls the local
response rate, with its inverse representing effective
damping.
The model parameters are therefore phenomenological
descriptors rather than quantities calibrated to particular
neurons or synapses.

The competition between the sinusoidal restoring force
and external driving produces the active-rotator dynamics
of oscillatory or excitable units
\cite{shinomoto_kuramoto_active_rotator}.
For an isolated, noiseless oscillator, the restoring
strength $W_i$ determines the tendency toward phase pinning,
allowing pinned or continuously rotating regimes depending
on the drive.
In the absence of driving, coupling, and noise, the
preferred resting phase is $\phi_i=A_i\Phi$ modulo $2\pi$.
The phase bias $\Phi$ shifts this reference position,
while $A_i$ determines the unit's sensitivity to both
external controls.

More generally, phase-oscillator models describe collective
behavior under disorder, noise, and heterogeneous coupling,
and support network inference from oscillatory signals
\cite{daido_disorder_oscillators, goldobin_pikovsky_noise, xu_heterogeneous_phase_oscillators, matsuki_kori_network_inference}.

Coupled nonlinear oscillators also provide physical
substrates for information processing and learning.
Their collective nonlinear dynamics and synchronization
have been exploited in neuromorphic architectures and
physical reservoir computing
~\cite{tanaka_physical_reservoir,
	romera_coupled_sto,kanao_sto_reservoir}.
This provides an additional motivation for inverse
response matching: in such systems, the relevant objective
may be to reproduce a desired collective input--output
behavior rather than to identify a unique microscopic
realization.

	\subsection{Statistical ensemble and disorder parameterization}

	Each realization of the oscillator network is specified by a set of
	node-wise intrinsic parameters and pairwise interaction coefficients.
	At the microscopic level, oscillator $i$ is characterized by three
	local parameters: the restoring strength $W_i>0$, the mobility
	$\alpha_i>0$, and the bias sensitivity $A_i>0$. Pairwise interactions are
	described by the coupling strengths $K_{ij}$ between oscillators $i$
	and $j$.

	In realistic disordered systems these quantities vary across nodes and pairwise interactions.
	Rather than fixing a single deterministic set of parameters, we
	therefore describe the system through a statistical ensemble of
	oscillator networks.

	The local parameters
	\begin{equation}
		\{W_i,\alpha_i,A_i\}
	\end{equation}
	are modeled as positive random variables drawn from Gamma
	distributions, while the pairwise couplings are sampled from a
	Gaussian ensemble
	\begin{equation}
		K_{ij}=K_{ji},
		\quad
		K_{ii}=0,
		\quad
		K_{ij}\sim\mathcal N(\mu_K,\sigma_K^2)
		\quad (i<j).
		\label{eq:interaction_sampling}
	\end{equation}
	Thus, individual realizations are microscopically disordered, while the
	network remains statistically homogeneous at the ensemble level.
	The full disorder ensemble may in general depend on many statistical
	quantities, including means, variances, and correlations of the microscopic parameters.

	In this work, the inverse problem is defined over the
	four-dimensional ensemble-parameter vector
	\begin{equation}
		\boldsymbol{\theta}=
		\big(
		\mu_W,\,
		\mu_K,\,
		\mu_A,\,
		\mathrm{CV}_K
		\big),
		\label{eq:theta_definition}
	\end{equation}
	where $\mu_W$ is the mean restoring strength, $\mu_K$ the mean pairwise
	coupling, $\mu_A$ the mean bias scale, and
	\[
	\mathrm{CV}_K=\frac{\sigma_K}{\mu_K}
	\]
	is the coefficient of variation of the coupling distribution. The parameter dimension is therefore
	\begin{equation}
		d_\theta\equiv\dim(\boldsymbol{\theta})=4.
	\end{equation} 
	 We assume that the characteristic scales of
	$\mu_W$, $\mu_K$, and $\mu_A$ are known approximately from calibration
	or prior physical knowledge, while their precise values remain unknown. Choosing units in which the nominal scales of the three mean parameters are
	of order unity, we allow $\mu_W$, $\mu_K$, and $\mu_A$ to vary from
	$0.5$ to $2$, spanning values from one half to twice their nominal scales.
	For the dimensionless coupling heterogeneity, we impose the analogous
	order-one range
	$0\leq\mathrm{CV}_K\leq1$, corresponding to fluctuations whose standard
	deviation does not exceed the mean coupling strength. The inverse search
	domain is therefore
	\begin{equation}
		0.5 \leq \mu_W,\mu_K,\mu_A \leq 2,
		\qquad
		0 \leq \mathrm{CV}_K \leq 1.
		\label{eq:inverse_bounds}
	\end{equation}
	These four quantities specify the disorder statistics varied in the inverse
	problem, while the remaining ensemble properties are held fixed as described
	below.

	For the numerical experiments considered here, the ensemble-mean mobility is
	fixed at $\mu_\alpha=1$ and is not included among the inferred parameters.
	Individual mobilities $\alpha_i$ nevertheless remain Gamma-distributed with
	finite variance. Likewise, the distribution widths of the local parameters
	that are not inferred are kept finite:
	\begin{equation}
		\mu_\alpha = 1,
		\qquad
		\mathrm{CV}_W
		=
		\mathrm{CV}_\alpha
		=
		\mathrm{CV}_A
		=
		0.25,
		\label{eq:fixed_ensemble_parameters}
	\end{equation}
	where $\mathrm{CV}$ is the coefficient of variation of the corresponding parameter. Thus, only $\mathrm{CV}_K$ is varied as part of the inferred parameter vector,
	while the other disorder channels retain finite quenched heterogeneity. No
	cross-parameter correlations (e.g., between $A$ and $K$) are imposed in the
	disorder ensemble used here, and self-coupling and inertial effects are absent.
	The microscopic parameters and their ensemble descriptions are summarized in
	Table~\ref{tab:parameter_summary}.

	\begin{table}[t]
		\caption{
			Microscopic parameters and ensemble models used in the numerical experiments.
			Local parameters are sampled from Gamma distributions, ensuring positivity,
			while pairwise couplings are drawn from a Gaussian ensemble. The inferred
			ensemble quantities are collected in the parameter vector
			$\boldsymbol{\theta}$ of Eq.~\eqref{eq:theta_definition}, while the fixed
			ensemble mean and coefficients of variation are given in
			Eq.~\eqref{eq:fixed_ensemble_parameters}.
		}
		\label{tab:parameter_summary}
		\begin{ruledtabular}
			\begin{tabular}{lcc}
				\textbf{Parameter} & \textbf{Symbol} & \textbf{Ensemble model} \\ \hline
				Local restoring weight & $W_i$ & Gamma \\
				Mobility & $\alpha_i$ & Gamma \\
				Bias scale & $A_i$ & Gamma \\
				Pairwise coupling & $K_{ij}$ & $\mathcal{N}(\mu_K,\sigma_K^2)$
			\end{tabular}
		\end{ruledtabular}
	\end{table}

	\begin{table}[t]
		\centering
		\caption{Default forward-simulation settings used in the numerical
			experiments.
		}
		\label{tab:simulation_parameters}
		\begin{tabular}{lc}
			\hline
			Parameter & Value \\
			\hline
			Network size, $N$ & $10$ \\
			Driving-force points, $N_F$ & $5$ \\
			{Phase-bias settings}, $N_\Phi$ & $2$ \\
			Observables per bias point & $2$ \\
			Response dimension, $d_y=2N_FN_\Phi$ & $20$ \\
			Time step, $\Delta\tau$ & $0.1$ \\
			Burn-in length & $100$ time steps \\
			Observation length & $900$ time steps \\
			Total integration length & $10^3$ time steps \\
			Disorder realizations & $80$ \\
			Stochastic repeats per realization & $2$ \\
			Effective noise intensity, $T$ & $0.15$ \\
			\hline
		\end{tabular}
	\end{table}

	\subsection{Multi-bias coarse observables and forward map}

	The experimentally relevant outputs of the network are not the full microscopic trajectories $\phi_i(\tau)$, but coarse observables obtained after averaging over time and disorder realizations. This reflects the fact that, in many physical systems, only macroscopic response functions are directly accessible. The inverse problem considered here is therefore formulated in terms of these coarse observables rather than the underlying microscopic dynamics.

	For a given realization of the network and fixed external bias parameters $(F,\Phi)$, the stochastic dynamics of Eq.~\eqref{eq:eom_overdamped} are integrated numerically using an Euler--Maruyama scheme (see Table~\ref{tab:simulation_parameters}  for simulation settings), after which the macroscopic observables are subsequently computed.

	From the simulated trajectories we construct two coarse observables.
	The first is the long-time averaged angular velocity
	\begin{equation}
		\bar{\omega} =
		\frac{1}{N}\sum_{i=1}^{N}
		\frac{\phi_i(\tau_f)-\phi_i(\tau_0)}{\tau_f-\tau_0},
		\label{eq:mean_velocity}
	\end{equation}
	which measures the mean rotation rate of the network. Here $[\tau_0,\tau_f]$ denotes the post-transient observation window following  the burn-in period, and the phase differences are computed using unwrapped phases.

	The second observable is the Kuramoto order parameter,
	\begin{equation}
		Z(\tau)=\frac{1}{N}\sum_{i=1}^{N}e^{i\phi_i(\tau)},
	\end{equation}
	which quantifies the instantaneous degree of phase synchronization across the network. We characterize synchronization through the time-averaged magnitude
	\begin{equation}
		M =
		\frac{1}{\tau_f-\tau_0}
		\int_{\tau_0}^{\tau_f}
		|Z(\tau)|\: d\tau,
		\label{eq:order_parameter}
	\end{equation}
	where the integral is taken over the same post-burn-in observation window.
	In numerical simulations, the integral is approximated by a discrete average over uniformly sampled times (see Table~\ref{tab:simulation_parameters}).

	For fixed ensemble parameters $\boldsymbol{\theta}$, effective noise intensity $T$, defined in Table~\ref{tab:simulation_parameters}, different disorder realizations and noise realizations generally produce different values of $\bar{\omega}$ and $M$. The macroscopic response is therefore characterized by an ensemble average over disorder realizations and independent stochastic repeats, denoted by $\langle\cdot\rangle$. This ensemble average is distinct from the time averages already appearing in the definitions of $\bar{\omega}$ and $M$.

	For a given ensemble parameter vector $\boldsymbol{\theta}$, the
	simulator is evaluated at a fixed effective noise intensity over a discrete set of
	bias conditions $(F_v,\Phi_u)$. We define the underlying macroscopic
	response as the ensemble expectation of the coarse observables,
	\begin{equation}
		\mathbf{y}(\boldsymbol{\theta})
		=
		\big(
		\langle \bar{\omega} \rangle,
		\langle M \rangle
		\big)_{u,v}
		\in \mathbb{R}^{d_y}.
		\label{eq:forward_map}
	\end{equation}
	In the numerical experiments, these ensemble expectations are
	approximated by finite averages over disorder realizations and
	independent stochastic repeats.	The response dimension is
	\begin{equation}
		d_y = 2 N_F N_\Phi,
	\end{equation}
	where $N_F$ and $N_\Phi$ denote the number of sampled driving-force and
	phase-bias values, respectively. Thus, two observables,
	$\langle\bar{\omega}\rangle$ and $\langle M\rangle$, are associated with
	each bias condition.
	{The five driving-force values are $F\in\{0,1.25,2.5,3.75,5\}$,
	and the two phase-bias values are $\Phi\in\{0,\pi\}$. The response vector
	contains all ten angular-velocity components followed by all ten
	order-parameter components, with conditions ordered first by driving force
	and then by phase bias ($\Phi=0$ before $\Phi=\pi$).}

	Throughout this work, the effective noise intensity is fixed at $T=0.15$,
	and its dependence is therefore suppressed in the notation. $T$ sets the strength of stochastic fluctuations,
	while the numbers of disorder realizations and independent stochastic repeats
	determine how accurately the resulting distribution of macroscopic responses
	is sampled. For a controlled comparison, $T$ and the averaging protocol are
	held fixed across all methods considered below. A systematic study of
	inverse-search performance as a function of effective noise intensity, disorder averaging,
	and stochastic-repeat count could probe the interplay between intrinsic
	response fluctuations, finite-sampling noise, and simulator cost. This
	noise--cost tradeoff lies beyond the present fixed-protocol benchmark.

	The function $\mathbf{y}$ defines the underlying forward map
	\begin{equation}
		\boldsymbol{\theta}
		\;\xrightarrow{\mathrm{forward}}\;
		\mathbf{y}(\boldsymbol{\theta}).
	\end{equation}
	Thus, the forward map associates the ensemble parameter vector
	$\boldsymbol{\theta}$ with a multi-component macroscopic response
	signature. In practice, the stochastic simulator provides the finite-sample estimate
	$\mathbf y(\boldsymbol{\theta})$ of this response.

	We illustrate the forward map in Fig.~\ref{fig:forward_response}, which shows the bias-dependent observables produced by the simulator for an illustrative ensemble-parameter vector. The resulting response signature provides an illustrative target for the
	inverse problem considered in the next section.

	\section{Response-Matching Inverse Problem}
	\label{sec:inverse}

	The forward map introduced in the previous section associates an
	ensemble-parameter vector $\boldsymbol{\theta}$ with a macroscopic
	response signature $\mathbf{y}(\boldsymbol{\theta})$. Given a prescribed
	target response $\mathbf{y}^\star$, the inverse task is to identify an
	admissible ensemble-parameter configuration whose simulated response
	reproduces that target.

	The target response $\mathbf{y}^\star$ is a finite-dimensional vector
	constructed from the same coarse observables evaluated at the discrete
	bias conditions that define the forward response, as illustrated, for example,
	in Fig.~\ref{fig:forward_response}. To assess performance across the inverse
	problem rather than for a single illustrative instance, we repeat the
	procedure over a cohort of independently generated target responses;
	the benchmark construction is described in Sec.~\ref{sec:benchmark_targets}.

	Formally, the inverse task is to find
	\begin{equation}
		\hat{\boldsymbol{\theta}} \in \Theta
		\quad \text{such that} \quad
		\mathbf{y}(\hat{\boldsymbol{\theta}})
		\approx
		\mathbf{y}^\star,
	\end{equation}
	where $\Theta\subset\mathbb{R}^{4}$ denotes the admissible
	ensemble-parameter domain. 	Fig.~\ref{fig:inverse_problem_schematic} summarizes this
	simulator-mediated inverse problem: the physical target and the stochastic
	simulator are subjected to the same controlled bias protocol, while the
	ensemble parameters $\boldsymbol{\theta}$ are varied so that the simulated
	response approaches the prescribed target response.

	The inverse problem is challenging for three related reasons. First,
	each evaluation of the forward map is computationally costly because it
	requires stochastic simulation and finite averaging over disorder
	realizations and independent stochastic repeats. Second, the dependence
	of the macroscopic response on the ensemble parameters is nonlinear and
	accessible only through numerical simulation. Third, inversion from the
	finite set of coarse macroscopic observables may be nonunique.

	To quantify agreement with the target response, we use the physical
	response mean absolute error
	\begin{equation}
		\label{eq:target_error}
		E(\boldsymbol{\theta})
		=
		\frac{1}{d_y}
		\left\|
		\mathbf{y}(\boldsymbol{\theta})
		-
		\mathbf{y}^\star
		\right\|_1
		=
		\frac{1}{d_y}
		\sum_{p=1}^{d_y}
		\left|
		y^p(\boldsymbol{\theta})-y_\star^p
		\right|.
	\end{equation}
	This quantity measures response-reproduction accuracy directly in the
	physical output coordinates.

	A small response error does not, in general, imply unique recovery of
	the parameter vector that generated the target. Because the measured
	observables are coarse and finite-dimensional, distinct
	ensemble-parameter configurations may produce similar macroscopic
	response signatures over the operating conditions of interest.
	Accordingly, response reproduction and parameter identification are
	treated as related but distinct inverse objectives in what follows.

	The difficulty of the inverse problem is not determined directly by the
	number of oscillators $N$. Increasing $N$ primarily increases the cost of
	each forward simulation, while the inverse search remains posed in terms of
	the ensemble-parameter vector $\boldsymbol{\theta}$ and the macroscopic
	response vector $\mathbf y$, rather than the microscopic state of an
	individual network. Its difficulty is therefore governed more directly by
	$\dim(\boldsymbol{\theta})$, $\dim(\mathbf y)$, and by the geometry,
	nonlinearity, and conditioning of the map between them. In fact, smaller
	networks can be more difficult to invert because finite-size fluctuations may
	obscure the dependence of the measured response on the underlying ensemble
	statistics, even though larger $N$ makes each forward evaluation more
	expensive.

	Simulator-mediated response-matching inverse problems of this kind may be
	approached from several directions. Bayesian inverse methods seek a posterior
	distribution over parameters conditioned on observations
	~\cite{stuart_bayesian_inverse}, with Gaussian-process constructions providing
	one route to parameter learning and uncertainty quantification in nonlinear
	dynamical systems~\cite{ye_guo_gp_dynamics,li_pigp_inverse}. Likelihood-free and
	simulation-based inference methods extend this viewpoint to settings where an
	explicit likelihood is unavailable
	~\cite{cranmer_sbi_review,papamakarios_snl}, with related neural approaches
	incorporating physical constraints directly into generative inverse models
	~\cite{guan_physics_informed_inverse}. Recent approaches also address
	high-dimensional black-box physical inference through ensemble message
	passing~\cite{mackinlay_genbp}, learned forward operators that are subsequently
	inverted against observations~\cite{mackinlay_fno_inversion}, and variational
	inference over transdimensional model and parameter spaces
	~\cite{davies_vti}. Alternatively, when the primary objective is direct
	reproduction of a target response under a restricted simulator budget, the
	target mismatch may be treated as an expensive black-box objective and
	addressed using Bayesian optimization or related sequential-design methods
	~\cite{brochu_bo_review,srinivas_gpucb}. Nonadaptive space-filling designs
	provide a complementary surrogate-free strategy for exploring the admissible
	parameter domain.
	
	These approaches differ in the quantity they seek to infer and in how
	they allocate expensive forward evaluations. Closely related inverse problems
	have also been studied directly in coupled-oscillator systems. Kato
	\textit{et al.} infer coupling strength and heterogeneity from the time series
	of a macroscopic Kuramoto order parameter using Bayesian parameter
	estimation~\cite{kato_kori_macroscopic_inference}, while Matsuki
	\textit{et al.} develop network-inference methods for recovering oscillator
	interactions from oscillatory signals in both synchronized and desynchronized
	regimes~\cite{matsuki_kori_network_inference}.

	\begin{figure}[t]
		\centering
		\begin{tikzpicture}[scale=1.05,>=latex]

			% Axes
			\draw[->] (0,0) -- (6.2,0) node[right] {$y_1$};
			\draw[->] (0,0) -- (0,5.6) node[above] {$y_2$};

			% Posterior mean and compatibility radius
			\coordinate (mu) at (2.5,2.9);
			\def\r{1.45}

			% Compatibility region
			\fill[blue!8] (mu) circle (\r);
			\draw[thick,blue] (mu) circle (\r);

			% Posterior mean
			\filldraw[blue] (mu) circle (2pt);
			\node[xshift=5pt,yshift=-7pt] at (mu)
			{$\boldsymbol{\mu}_t(\boldsymbol{\theta})$};

			% Target: moved to upper-left sector
			\coordinate (yt) at (1.65,3.70);
			\filldraw[red] (yt) circle (2pt);
			\node[xshift=-2pt,yshift=2pt] at (yt) {$\mathbf y^\star$};

			% Residual vector
			\draw[->,thick] (mu) -- (yt);
			\node[left] at (2.05,3.25)
			{$\mathbf r_t(\boldsymbol{\theta})$};

			% Compatibility radius
			\draw[densely dotted] (2.5,2.9) -- (3.61,3.83);

			\node[blue] at (3.00,3.7)
			{$\rho_t(\boldsymbol{\theta})$};

			\node[blue,align=left] at (4.0,4.7)
			{$\rho_t(\boldsymbol{\theta})=
				\beta\sqrt{\lambda_{\max}
					[\boldsymbol{\Sigma}_t(\boldsymbol{\theta})]}$};

			% Barrier cue on lower-right boundary
			\draw[very thick,orange!80!black]
			(3.95,3.03)
			arc[start angle=5,end angle=45,radius=1.45];

			\node[orange!80!black,align=center] at (5.0,2.95)
			{\small barrier grows\\[-1mm]
				\small near boundary};

			% Compatibility condition: keep fixed at bottom
			\node[align=center] at (2.35,0.40)
			{\small compatible if\\[-1mm]
				\small $\|\mathbf r_t(\boldsymbol{\theta})\|_2
				\leq \rho_t(\boldsymbol{\theta})$};

		\end{tikzpicture}

		\caption{Schematic response-space view of the MUTACO compatibility test for a
			candidate parameter vector $\boldsymbol{\theta}$. The axes \(y_1\) and \(y_2\) denote two representative components of the multicomponent response vector. The surrogate posterior mean
			response $\boldsymbol{\mu}_t(\boldsymbol{\theta})$ is surrounded by an
			uncertainty-dependent compatibility region. The target is treated as compatible when it lies within this region; the
			barrier diverges logarithmically,
			$\sim-\log(1-r/\rho)$, as the compatibility boundary
			$r\to\rho^{-}$ is approached from the interior.}
		\label{fig:mutaco_barrier_exploration_schematic}
	\end{figure}
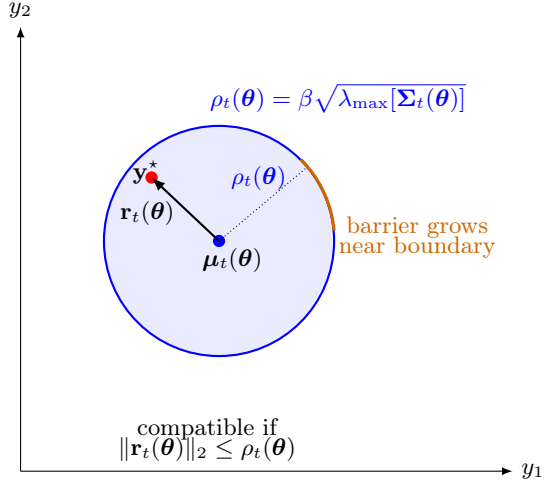

	\begin{figure}[t]
		\centering
		\resizebox{1\columnwidth}{!}{
			\begin{tikzpicture}[
				node distance=0.75cm,
				box/.style={
					draw,
					fill=gray!10,
					rectangle,
					rounded corners=1.5pt,
					minimum height=0.75cm,
					minimum width=0.6\columnwidth,
					align=center,
					font=\small
				},
				smallbox/.style={
					draw,
					fill=gray!10,
					rectangle,
					rounded corners=1.5pt,
					minimum height=0.75cm,
					minimum width=0.38\columnwidth,
					align=center,
					font=\small
				},
				decision/.style={
					draw,
					fill=gray!10,
					diamond,
					aspect=2.2,
					inner sep=1.5pt,
					minimum width=0.72\columnwidth,
					align=center,
					font=\small
				},
				arrow/.style={->, thick}
				]

				\node[box] (initial) {Initial simulator evaluations};

				\node[box, below=of initial] (gp)
				{Fit GP surrogate};

				\node[decision, below=0.85cm of gp] (compat)
				{Compatible with\\target response $y_\star$?};

				\node[smallbox, below left=1.0cm and 0.05cm of compat] (barrier)
				{Barrier\\refinement};

				\node[smallbox, below right=1.0cm and 0.05cm of compat] (explore)
				{Uncertainty-driven\\exploration};

				\node[box, below=2.0cm of compat] (next)
				{Next candidate design};

				\node[box, below=of next] (update)
				{Update data and refit GP};

				\node[font=\small, below=0.45cm of update] (repeat)
				{Repeat until evaluation budget is exhausted};

				\node[box, below=0.65cm of repeat] (recommend)
				{Recommend $\hat{\boldsymbol{\theta}}$};

				\draw[arrow] (initial) -- (gp);
				\draw[arrow] (gp) -- (compat);

				\draw[arrow] (compat.west) -|
				node[pos=0.25, above, font=\small] {Yes}
				(barrier.north);

				\draw[arrow] (compat.east) -|
				node[pos=0.25, above, font=\small] {No}
				(explore.north);

				\draw[arrow] (barrier.south) |- (next.west);
				\draw[arrow] (explore.south) |- (next.east);

				\draw[arrow] (next) --
				node[midway, right, font=\small, align=left]
				{evaluate stochastic\\simulator}
				(update);

				\draw[arrow] (update) -- (repeat);
				\draw[arrow] (repeat) -- (recommend);

			\end{tikzpicture}
		}
		\caption{
			Flowchart of the MUTACO target-matching search.
			After an initial design is used to fit the GP surrogate, sequential
			candidates are selected through either compatibility-guided barrier
			refinement or uncertainty-driven exploration. Each selected candidate
			is evaluated with the stochastic simulator and the surrogate is updated
			until the evaluation budget is exhausted, after which the final
			parameter configuration $\hat{\boldsymbol{\theta}}$ is recommended.
			See the main text and Appendix~\ref{app:mutaco_algorithm} for details.
		}
		\label{fig:mutaco_flowchart}
	\end{figure}
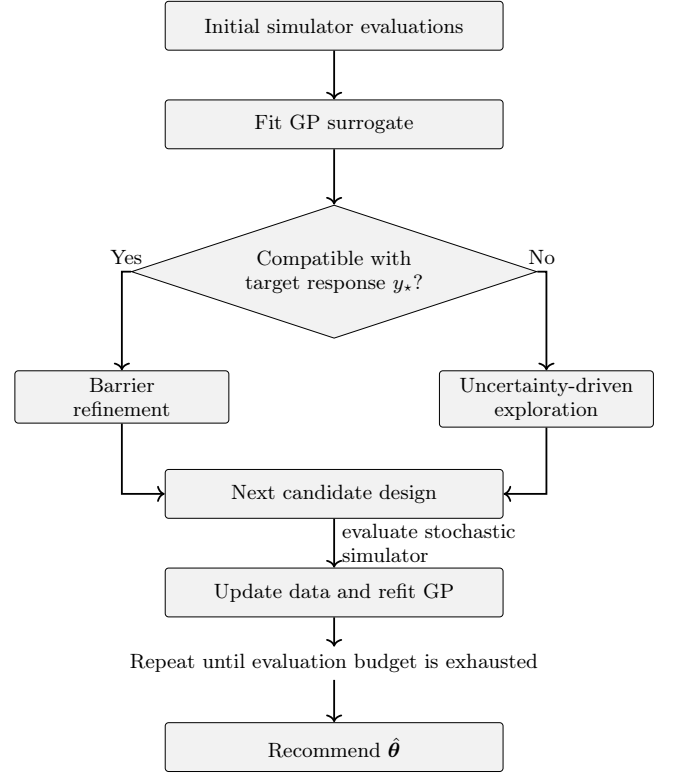

The oscillator-focused studies above emphasize recovery of underlying
model parameters or interaction structure from observed dynamics. The problem
considered here is different: the primary objective is not posterior
identification of the generating model, but simulator-efficient search for
ensemble parameters whose macroscopic response matches a prescribed target
under a restricted forward-simulation budget. This response-matching
formulation motivates the sequential design developed in the following
section.

	\section{MUTACO: Compatibility-Aware Sequential Design}
	\label{sec:mutaco}

	\subsection{Gaussian-process response model}
	\label{sec:mutaco_gp}

	Evaluating the stochastic simulator throughout the parameter domain
	$\Theta$ would in principle determine the full response map
	$\mathbf y(\boldsymbol{\theta})$, but doing so is prohibitively costly.
	MUTACO therefore represents the unknown response away from the simulated
	parameter configurations by a probabilistic model. Specifically, we use a
	correlated multitask Gaussian process
	\cite{rasmussen_williams_gp,alvarez_multitask_gp,bonilla_multitask_gp,chowdhury_multitask_gp},
	\begin{equation}
		\boldsymbol{\theta}
		\longmapsto
		\mathbf f(\boldsymbol{\theta})
		\sim
		\mathcal{GP}\!\left[
		\mathbf m(\boldsymbol{\theta}),
		\mathbf K(\boldsymbol{\theta},\boldsymbol{\theta}')
		\right],
		\label{eq:main_gp_field}
	\end{equation}
	where $\mathbf f(\boldsymbol{\theta})\in\mathbb{R}^{d_y}$ denotes the latent
	ensemble-averaged macroscopic response, $\mathbf m$ its prior mean, and
	$\mathbf K$ its matrix-valued covariance kernel. This Gaussian-process model
	serves as a surrogate for the response map at parameter configurations that
	have not yet been simulated. For completeness, a compact derivation of the standard
	Gaussian-process conditioning relations, together with their Gaussian-field
	interpretation, is given in Appendix~\ref{app:gp_theory}.

	The elements of the covariance kernel are
	\begin{equation}
		K^{pp'}(\boldsymbol{\theta},\boldsymbol{\theta}')
		=
		\operatorname{Cov}
		\!\left[
		f^p(\boldsymbol{\theta}),
		f^{p'}(\boldsymbol{\theta}')
		\right],
		\label{eq:main_gp_kernel}
	\end{equation}
	so that $K^{pp'}$ gives the covariance between response component $p$ at
	$\boldsymbol{\theta}$ and response component $p'$ at
	$\boldsymbol{\theta}'$. The kernel therefore encodes correlations both
	across the ensemble-parameter domain and among the different components of
	the macroscopic response.

	Suppose that the simulator has been evaluated at parameter vectors
	$\boldsymbol{\theta}_1,\ldots,\boldsymbol{\theta}_t$, producing finite-sample
	response estimates $\mathbf y_1,\ldots,\mathbf y_t$. The accumulated data are
	then
	\begin{equation}
		\mathcal D_t
		=
		\left\{
		\bigl(\boldsymbol{\theta}_i,\mathbf y_i\bigr)
		\right\}_{i=1}^{t}.
		\label{eq:main_gp_data}
	\end{equation}
	Conditioning the Gaussian process on $\mathcal D_t$ gives, at any
	candidate $\boldsymbol{\theta}$,
	\begin{equation}
		\mathbf f(\boldsymbol{\theta})\mid\mathcal D_t
		\sim
		\mathcal N\!\left[
		\boldsymbol{\mu}_t(\boldsymbol{\theta}),
		\boldsymbol{\Sigma}_t(\boldsymbol{\theta})
		\right].
		\label{eq:main_gp_posterior}
	\end{equation}
	The posterior mean
	$\boldsymbol{\mu}_t(\boldsymbol{\theta})$ is the current prediction of the
	latent macroscopic response, while
	$\boldsymbol{\Sigma}_t(\boldsymbol{\theta})$ quantifies the remaining
	uncertainty in that response and its correlations across response components.
	This is uncertainty in the inferred response map, rather than the fluctuations from stochastic forcing, disorder, or finite sampling entering an individual simulator evaluation. The observation-noise model and the explicit multitask
	conditioning formulas used to obtain
	$\boldsymbol{\mu}_t$ and $\boldsymbol{\Sigma}_t$ are given in
	Appendix~\ref{app:gp}.

	Rather than reducing the multicomponent response immediately to a single
	scalar mismatch, such as the response error
	$E(\boldsymbol{\theta})$ defined in Eq.~\eqref{eq:target_error}, MUTACO
	retains both the response-vector structure and its correlated posterior
	uncertainty when selecting the next simulator evaluation. This is related in
	spirit to multiobjective Bayesian optimization, where multiple response
	components are retained rather than collapsed a priori to a single objective
	~\cite{knowles_multiobjective_bo}, although here the goal is target
	compatibility rather than Pareto optimization. This distinguishes MUTACO from
	the scalar qLogNEI benchmark described in Appendix~\ref{app:qlognei}, which
	performs acquisition directly on the scalar response-error objective.

	The Gaussian-process construction also admits a useful interpretation as a
	Gaussian field theory, a standard framework in statistical physics and quantum
	field theory~\cite{kardar_statistical_fields}. In this language, the covariance
	kernel $\mathbf K$ plays the role of a two-point correlation function, or
	propagator, while $\mathbf K^{-1}$ defines the corresponding quadratic
	fluctuation operator. The analogy is mathematical rather than dynamical:
	$\boldsymbol{\theta}$ labels ensemble-parameter space rather than physical
	space or time. Appendix~\ref{app:gp_theory} develops this correspondence and
	the standard conditioning relations for a single response component, while
	Appendix~\ref{app:gp} gives their correlated multicomponent realization used
	by MUTACO.

	\subsection{Target compatibility and acquisition}
	\label{sec:mutaco_acquisition}

	For the compatibility construction, response quantities are expressed in the
	standardized coordinates defined in Appendix~\ref{app:gp_surrogate}; tildes
	are suppressed here for readability. MUTACO compares the displacement of the
	posterior mean response from the prescribed target,
	\begin{equation}
		\mathbf r_t(\boldsymbol{\theta})
		=
		\boldsymbol{\mu}_t(\boldsymbol{\theta})
		-
		\mathbf y^\star ,
		\label{eq:main_residual}
	\end{equation}
	with the remaining latent posterior uncertainty. The target is regarded as
	compatible with the current surrogate prediction when
	\begin{equation}
		\left\|
		\mathbf r_t(\boldsymbol{\theta})
		\right\|_2^2
		\leq
		\beta^2
		\lambda_{\max}
		\!\left[
		\boldsymbol{\Sigma}_t(\boldsymbol{\theta})
		\right].
		\label{eq:main_compatibility}
	\end{equation}
	Here $\lambda_{\max}$ denotes the largest eigenvalue of the posterior
	covariance and $\beta$ sets the {uncertainty scale} used in the compatibility
	criterion. Thus, a larger mismatch can remain admissible where the surrogate
	is uncertain, whereas the compatible region contracts as the response becomes
	more precisely determined. This construction is related to constrained Bayesian optimization, where
	probabilistic surrogate information is used to restrict search to admissible
	regions~\cite{gardner_constrained_bo}; here, however, admissibility is defined by an uncertainty-scaled compatibility
	criterion relative to the target response rather than by an external
	constraint function.

	Fig.~\ref{fig:mutaco_barrier_exploration_schematic}
	illustrates the response-space geometry of this criterion. The {choice of
	uncertainty scale} is given in Appendix~\ref{app:gp_acquisition}.

	Candidates identified within the strict interior of the compatibility region
	are refined toward the target using a logarithmic barrier that penalizes
	approach to the compatibility boundary. If the candidate search identifies no
	finite strict-interior compatible point, MUTACO instead directs the next
	simulation toward the parameter region with the largest latent posterior
	uncertainty,
	\begin{equation}
		\boldsymbol{\theta}_{t+1}
		\approx
		\arg\max_{\boldsymbol{\theta}\in\Theta}
		\lambda_{\max}
		\!\left[
		\boldsymbol{\Sigma}_t(\boldsymbol{\theta})
		\right].
		\label{eq:main_uncertainty_acquisition}
	\end{equation}
	Thus, sequential acquisition alternates between target-directed
	refinement when a compatible region has been identified and uncertainty-driven
	exploration otherwise.
	The resulting sequential logic is summarized in
	Fig.~\ref{fig:mutaco_flowchart}, while the complete numerical construction of
	both acquisition branches is given in Appendix~\ref{app:gp_acquisition}.

	After the simulator budget $B$ has been exhausted, MUTACO forms its
	final recommendation from the posterior mean of the final multitask GP. The
	recommended parameter vector is obtained by continuously minimizing the mean
	absolute error between the posterior-mean response and the target in physical
	response units,
	\begin{equation}
		\widehat{\boldsymbol{\theta}}
		\approx
		\arg\min_{\boldsymbol{\theta}\in\Theta}
		\frac{1}{d_y}
		\sum_{p=1}^{d_y}
		\left|
		\mu_B^{p,\mathrm{phys}}(\boldsymbol{\theta})
		-
		y_\star^p
		\right|.
		\label{eq:main_terminal}
	\end{equation}
	This continuous terminal optimization uses only the fitted surrogate and
	therefore requires no additional simulator evaluations. By contrast, an
	evaluated-point terminal policy restricts the final recommendation to parameter
	vectors that were explicitly simulated during the search. The continuous terminal recommendation reduces the mean held-out response
	error, as quantified in Appendix~\ref{app:terminal}. The
	continuous terminal procedure used by MUTACO is detailed in
	Appendix~\ref{app:gp_terminal}, and the complete algorithm is summarized in
	Fig.~\ref{fig:mutaco_algorithm}.

	\begin{table}[t]
		\centering
		\caption{
			Summary of the response-matching benchmark design.
		}
		\label{tab:experimental_design}
		\begin{ruledtabular}
			\begin{tabular}{lc}
				\textbf{Quantity} & \textbf{Value} \\
				\hline
				Target instances & $120$ \\
				Parameter dimension & $d_\theta=4$ \\
				Response dimension & $d_y=20$ \\
				Simulator budgets & $B=40,\ 100$ \\
				Initial design size & $n_0=16$ \\
				Sequential evaluations & $24,\ 84$ \\
				Response-matching methods & MUTACO, qLogNEI, Sobol
			\end{tabular}
		\end{ruledtabular}
	\end{table}

	\section{Experimental protocol}
	\label{sec:experiments}

	The numerical experiments are designed to evaluate MUTACO as a
	simulator-efficient response-matching method. We distinguish between
	two complementary questions. First, we compare MUTACO with alternative
	search strategies that solve the same response-matching problem under
	the same simulator budget. Second, we compare response matching with
	simulation-based parameter-inference methods, for which recovery of the
	generating parameter vector is a more natural objective.

	All methods are evaluated on the same collection of synthetic
	targets and within the same admissible parameter domain. Performance is
	assessed primarily through the accuracy with which the final recommended
	parameter configuration reproduces the prescribed macroscopic response.
	Parameter recovery is reported separately as a diagnostic of
	nonuniqueness rather than as the primary success criterion.	The main features of the experimental design are summarized in
	Table~\ref{tab:experimental_design}.

	\subsection{Benchmark targets}
	\label{sec:benchmark_targets}

	The benchmark consists of 120 {predefined} inverse-problem
	instances. For each instance, a generating parameter vector
	$\boldsymbol{\theta}^{\star}\in\Theta$ is selected and the forward
	simulator is used to construct a target response
	$\mathbf y^{\star}$. Once generated, each target is held fixed and reused
	across all methods and evaluation budgets.
	The 120 generating vectors are generated by a scrambled four-dimensional
	Sobol design over $\Theta=[0.5,2]^3\times[0,1]$.

	The generating vector $\boldsymbol{\theta}^{\star}$ is retained only for
	retrospective analysis. The response-matching methods are given
	$\mathbf y^{\star}$ but not $\boldsymbol{\theta}^{\star}$, and their
	objective is to find a parameter configuration
	$\hat{\boldsymbol{\theta}}$ whose simulated response reproduces the
	target. Consequently, accurate response matching does not require
	$\hat{\boldsymbol{\theta}}$ to recover
	$\boldsymbol{\theta}^{\star}$. The forward-simulation protocol used to construct targets and evaluate
	candidate parameter vectors is the same as that described in
	Sec.~\ref{sec:oscillator}.

	\subsection{Simulator budgets and accounting}
	\label{sec:budgets}

	Response-matching performance is evaluated under two simulator-call
	budgets,
	\begin{equation}
		B\in\{40,100\}.
	\end{equation}
	Each search begins from $n_0=16$ initial parameter points. The two budgets
	therefore correspond to 16 initial evaluations followed by 24 or 84
	additional evaluations, respectively.
	A simulator call corresponds to one evaluation of the stochastic
	forward map at a selected parameter vector, using the ensemble-averaging
	protocol defined in Sec.~\ref{sec:oscillator}. Only simulator
	evaluations performed while searching for a solution count toward the
	budget.

	Target construction is performed before the inverse search and is not
	charged to $B$. Likewise, optimization of a fitted surrogate without
	additional forward simulation does not consume simulator calls.
	Held-out assessment of the final recommendation is performed only after
	the search has terminated and is accounted for separately.

	The $B=40$ and $B=100$ experiments are treated as separate runs rather
	than as nested checkpoints of a common trajectory. Thus, each
	target--method--budget combination constitutes an independent inverse
	search. The complete benchmark protocol is summarized in
	Table~\ref{tab:benchmark_protocol}.

	\begin{figure*}[t]
		\centering
		\includegraphics[width=0.82\textwidth]
		{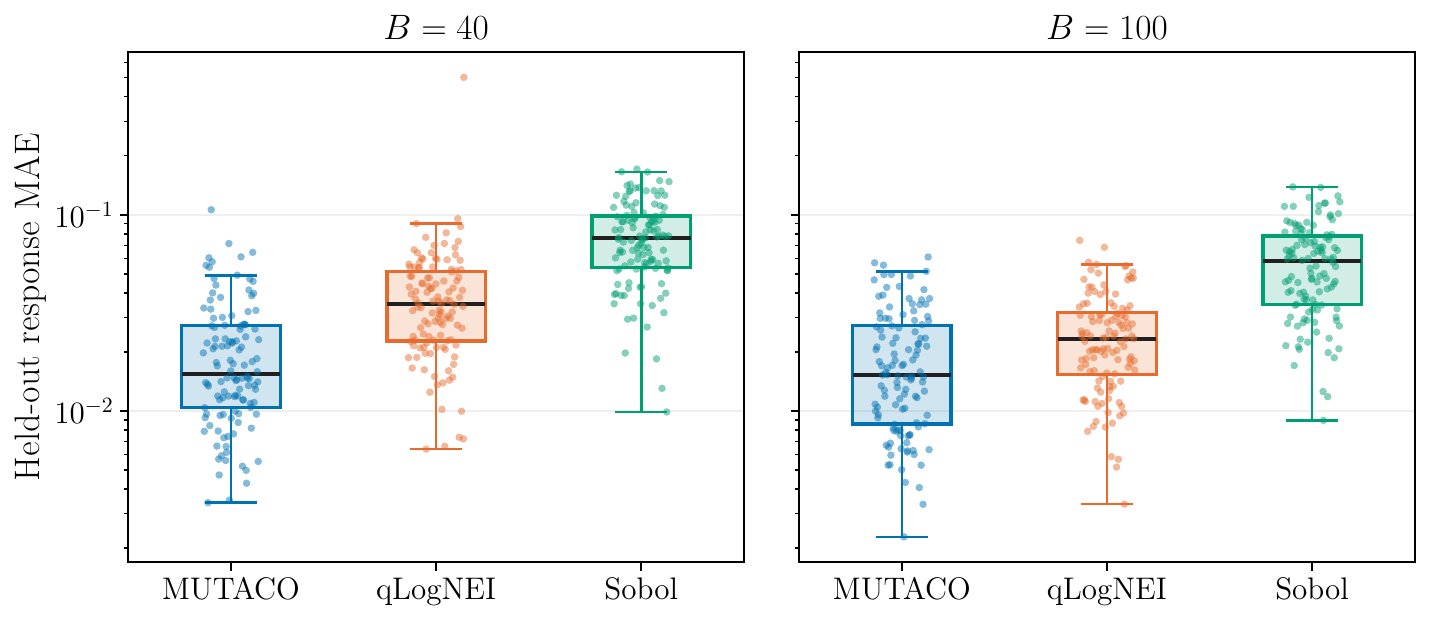}
		\caption{
			Held-out physical response error, defined in  Eq.~\eqref{eq:heldout_response_error}, across the 120-target benchmark cohort at
			simulation budgets $B=40$ and $B=100$. MUTACO is compared with scalar
			qLogNEI and nonadaptive scrambled-Sobol search. Each point corresponds to
			one of the 120 frozen target instances. The response error is the
			component-mean absolute difference between the held-out simulated response
			at the recommended parameter vector and the prescribed target response.
			Lower values indicate better response reconstruction. Boxes show the interquartile range with the median indicated by the central
			line. Whiskers extend to the most extreme observations within $1.5$ times
			the interquartile range, and all 120 observations, including points beyond
			the whiskers, are shown individually. Numerical
			means are reported in the text.
		}
		\label{fig:primary_response_mae}
	\end{figure*}
	\subsection{Compared methods}
	\label{sec:compared_methods}

	The primary comparison considers three response-matching strategies under the
	same simulator budget: MUTACO, scalar qLogNEI as an objective-aligned
	Bayesian-optimization baseline, and nonadaptive scrambled-Sobol search.
	Scalar qLogNEI directly models the physical target-matching error as a
    one-dimensional objective, whereas MUTACO models the multicomponent response
    and its correlated posterior uncertainty and uses these quantities in its
    compatibility-aware acquisition rule. The comparison therefore tests
    MUTACO against an objective-aligned scalarization of the same response-matching
    problem under the same simulator budget. Sobol provides a surrogate-free
	space-filling reference within the complete-procedure comparison.
	Numerical details of the
	qLogNEI and Sobol baselines are given in Appendix~\ref{app:baselines}.

	A separate comparison considers neural posterior estimation
	(NPE)~\cite{papamakarios_murray_npe,greenberg_snpe},
	sequential neural posterior estimation
	(SNPE)~\cite{greenberg_snpe},
	neural likelihood estimation
	(NLE)~\cite{papamakarios_snl},
	and neural ratio estimation
	(NRE)~\cite{hermans_nre, durkan_nre_b}.
	In the implementation used here, NPE is run in a single round, SNPE uses
	three rounds of SNPE-C/APT, NLE is single-round NLE-A, and NRE is
	single-round NRE-B/SRE.
	These methods address the related but distinct problem of inferring
	parameters from responses and are therefore treated as complementary rather
	than objective-matched baselines. Their recommendations are assessed using
	the same physical response and parameter-recovery metrics.

	\subsection{Randomization and held-out assessment}
	\label{sec:randomization}

	The first 16 parameter locations are common across MUTACO, scalar qLogNEI,
	and Sobol within each target--budget experiment, while simulator randomness
	during the search is generated independently for each method. After each
	method produces its final recommendation, the recommendations are evaluated
	under the same method-independent held-out stochastic context, so that the
	assessment is paired across methods. These held-out evaluations are performed
	only after the search has terminated and do not count toward the simulator
	budget $B$.
	Further details of the randomization and held-out assessment protocol are
	given in Appendix~\ref{app:numerics}. A complete summary of the benchmark
	design and simulator-call accounting is provided in
	Table~\ref{tab:benchmark_protocol}.

	\subsection{Evaluation metrics}
	\label{sec:evaluation_metrics}

	The primary endpoint metric is the held-out physical response error
	\begin{equation}
		E_y(\hat{\boldsymbol{\theta}})
		=
		\frac{1}{d_y}
		\left\|
		\mathbf y(\hat{\boldsymbol{\theta}})
		-
		\mathbf y^\star
		\right\|_1,
		\label{eq:heldout_response_error}
	\end{equation}
	where the response in Eq.~\eqref{eq:heldout_response_error} is obtained
	from the held-out simulator evaluation. This metric directly measures
	the physical objective of reproducing the target response.

	To distinguish response reconstruction from parameter identification,
	we separately report the distance between the recommended and generating
	parameter vectors,
	\begin{equation}
		E_\theta
		=
		\left\|
		\widehat{\boldsymbol{\theta}}
		-
		\boldsymbol{\theta}^{\star}
		\right\|_2 .
		\label{eq:parameter_error}
	\end{equation}
	The norm is evaluated directly in the four unnormalized physical parameter
	coordinates.
	This quantity is used as a diagnostic rather than as the optimization
	criterion for MUTACO.

	Unless stated otherwise, reported performance is aggregated over the same
	predefined set of 120 target instances.

	\begin{figure*}[t]
		\centering
		\includegraphics[width=0.82\textwidth]
		{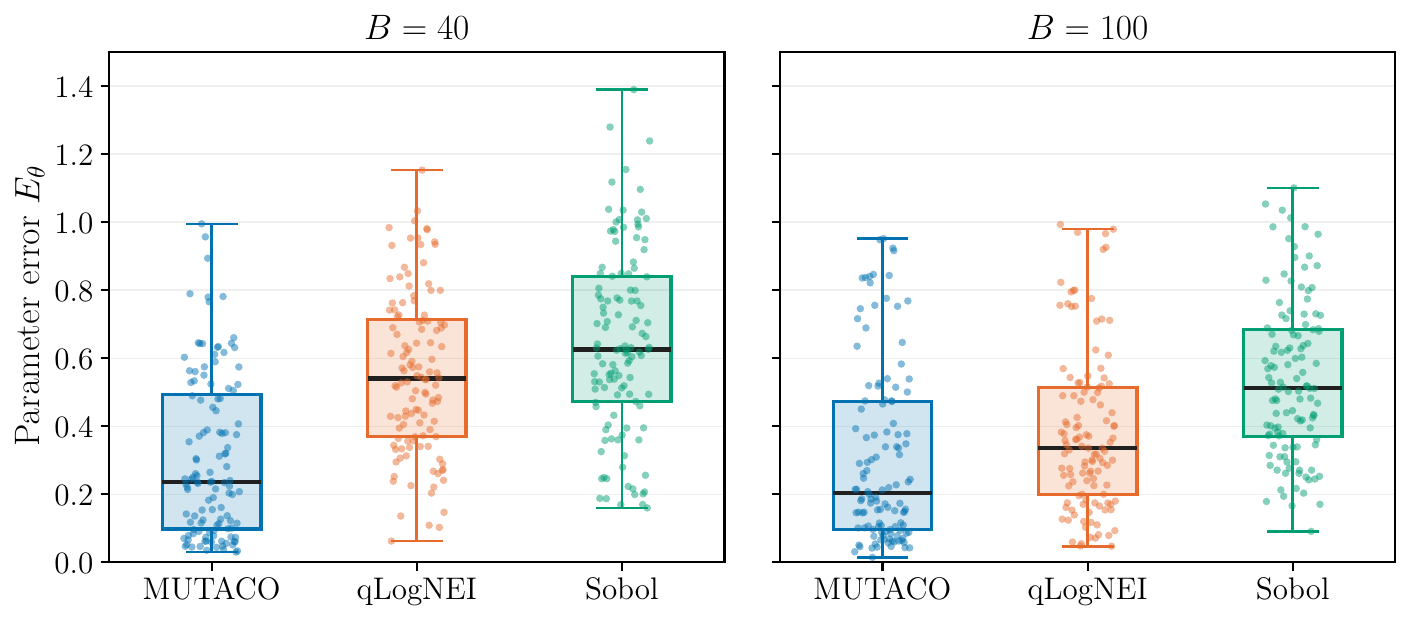}
		\caption{
			Parameter-recovery error, defined in Eq.~\ref{eq:parameter_error}, across the 120-target benchmark cohort at
			$B=40$ and $B=100$. The error is the Euclidean distance
			$\|\widehat{\boldsymbol{\theta}}-\boldsymbol{\theta}^{\star}\|_2$
			in the four unnormalized physical parameter coordinates. The generating
			parameter vector is used only for retrospective evaluation and is not
			available to the search methods. Lower values indicate more accurate
			recovery of the generating disorder parameters. See the caption of Fig.~\ref{fig:primary_response_mae} for the box plot convention. Numerical means are reported in the text. 
		}
		\label{fig:primary_parameter_error}
	\end{figure*}

	\section{Results}
	\label{sec:results}

	\subsection{Collective response and illustrative reconstruction}
	\label{sec:illustrative_reconstruction}

	We first illustrate the response-matching problem for the oscillator network
	before considering the full benchmark ensemble. Fig.~\ref{fig:forward_response}
	shows the macroscopic rotation and synchronization responses generated by one
	illustrative disorder ensemble. Both observables vary systematically with the
	external drive while retaining visible fluctuations across disorder
	realizations, emphasizing that the inverse problem concerns reproduction of a
	stochastic collective response rather than a deterministic trajectory.

	Fig.~\ref{fig:response_reconstruction} shows the corresponding MUTACO
	reconstruction under the $B=40$ simulation budget. The recommended parameter
	vector reproduces both the mean angular-velocity response and the
	time-averaged synchronization response across the sampled driving conditions,
	despite not coinciding exactly with the generating parameter vector. The
	reconstructed curves remain within the characteristic variability of the
	target ensemble over much of the response range, illustrating the distinction
	between reproducing the collective behavior and recovering the particular
generating ensemble-parameter vector.

\begin{figure*}[t]
		\centering
		\includegraphics[width=0.82\textwidth]
		{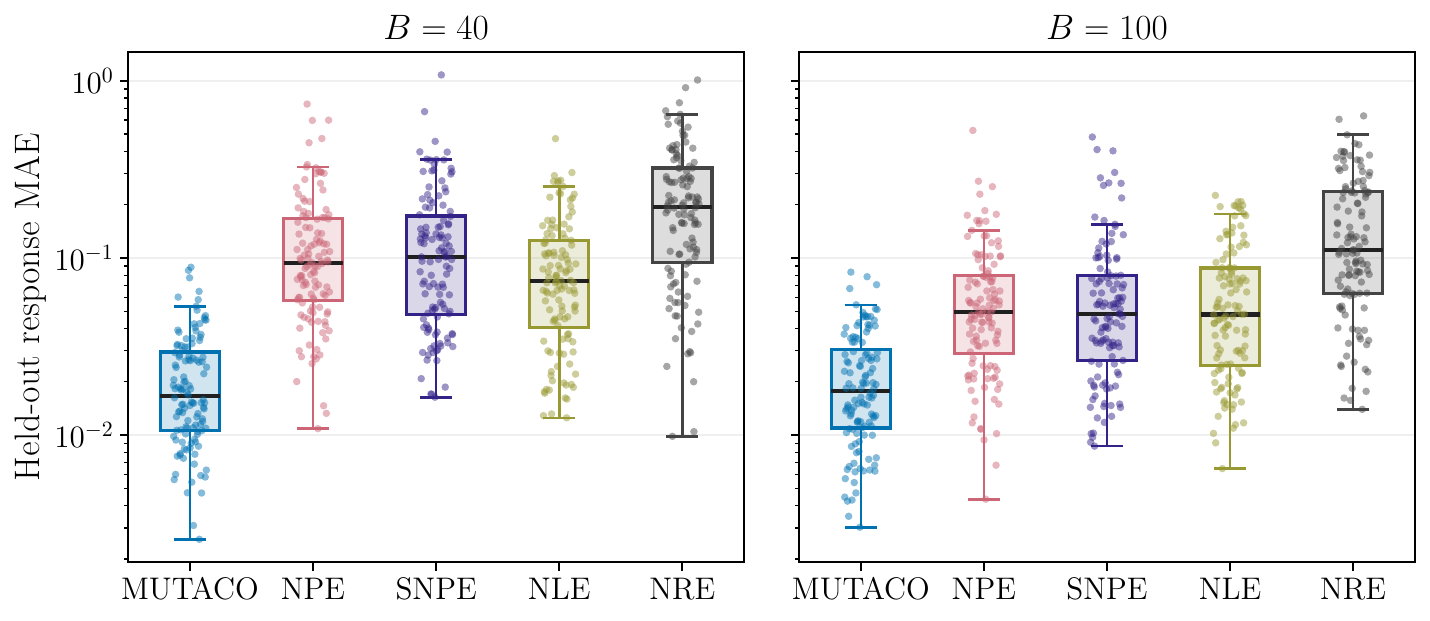}
		\caption{
			Held-out physical response error, Eq.~\eqref{eq:heldout_response_error}, for MUTACO and the simulation-based
			inference methods NPE, SNPE, NLE, and NRE at simulation budgets
			$B=40$ and $B=100$. All methods are assessed on the same target
			cohort using the component-mean absolute response error. The NPE/SNPE and
			NLE/NRE results originate from separately validated SBI benchmark
			campaigns and are combined here through their common assessment protocol.
			Lower values indicate better reproduction of the prescribed macroscopic
			response.  See the caption of Fig.~\ref{fig:primary_response_mae} for the box plot convention. Numerical means are reported in the text. 
		}
		\label{fig:sbi_response_mae}
	\end{figure*}

This example motivates the benchmark analysis below, where response fidelity
and parameter recovery are evaluated separately across 120 targets and
under fixed simulator budgets. The illustrative reconstruction demonstrates
the qualitative form of response equivalence, while the systematic
population-level benchmark analysis is presented in
Figs.~\ref{fig:primary_response_mae}--\ref{fig:sbi_response_parameter}.

	\subsection{Response matching under limited simulation budgets}
	\label{sec:primary_response}

	We next compare MUTACO with scalar qLogNEI and scrambled-Sobol search across
	the full 120-target benchmark. Fig.~\ref{fig:primary_response_mae} shows the
	held-out physical response error at simulation budgets $B=40$ and $B=100$.
	At both budgets, MUTACO yields the lowest error distribution among the three
	methods, with the largest separation occurring in the more restrictive
	$B=40$ regime.

At $B=40$, the mean held-out response error is $2.15\times10^{-2}$ for MUTACO,
compared with $4.15\times10^{-2}$ for scalar qLogNEI and $7.93\times10^{-2}$ for Sobol.
MUTACO outperforms scalar qLogNEI on 99 of the 120 paired targets. At
$B=100$, the corresponding mean errors are $1.96\times10^{-2}$,
$2.55\times10^{-2}$, and $5.91\times10^{-2}$, respectively, with MUTACO
outperforming scalar qLogNEI on 84 of 120 targets. Thus, increasing the simulation budget improves the
	objective-aligned qLogNEI baseline substantially, but MUTACO retains the
	lowest mean held-out response error at both budgets.

  {The complete MUTACO search-and-recommendation procedure yields lower
  held-out response errors than the nonadaptive Sobol procedure across this benchmark.}
  The comparison with scalar qLogNEI further shows that
  the complete scalar-objective qLogNEI procedure yields poorer held-out
  reconstruction than the multicomponent, compatibility-aware MUTACO procedure.
  The terminal recommendation in each method is chosen consistently with the
  information available from its search model as explained in Appendix~\ref{app:baselines}; the benchmark therefore compares
  complete search-and-recommendation procedures under the same simulator budget
  and common held-out response metric, rather than acquisition rules in
  isolation. The contribution of MUTACO's continuous terminal search is examined
  separately in Appendix~\ref{app:terminal}.
  The separation is largest under the tighter simulation budget, where the
  efficient use of predictive response structure and uncertainty is particularly
  consequential.

	\subsection{Recovery of the generating disorder parameters}
	\label{sec:primary_parameter}

	We next examine whether the response-matching advantage is accompanied by
	improved recovery of the generating ensemble parameters.
	Fig.~\ref{fig:primary_parameter_error} shows the parameter-recovery error
	$E_\theta$ of Eq.~\eqref{eq:parameter_error}, evaluated over the
	120-target benchmark cohort. Although parameter recovery is not the
	optimization objective of MUTACO, it nevertheless yields the lowest parameter
	error among the three primary search methods at both simulation budgets.

	At $B=40$, the mean parameter errors are $0.31$ for MUTACO, $0.56$ for
	scalar qLogNEI, and $0.65$ for Sobol, with corresponding medians of
	$0.24$, $0.54$, and $0.63$. At $B=100$, the mean errors are $0.31$,
	$0.38$, and $0.54$, respectively, while the medians are $0.204$, $0.34$,
	and $0.51$. Increasing the budget therefore substantially improves parameter
	recovery for qLogNEI and Sobol, whereas MUTACO already attains comparatively
	low parameter error in the $B=40$ regime.

	These results show that response-focused optimization does not require
	sacrificing parameter recovery in this benchmark. At the same time, the
	difference between the response-error and parameter-error distributions
	indicates that the two objectives are not equivalent: parameter vectors that
	are relatively distant from the generating vector can still reproduce the
	macroscopic response accurately. This distinction is examined more directly
	below in the comparison with simulation-based inference methods.

	\subsection{Comparison with simulation-based inference}
	\label{sec:sbi_comparison}

	We next compare MUTACO with four simulation-based inference methods: NPE,
	SNPE, NLE, and NRE. Fig.~\ref{fig:sbi_response_mae} compares the held-out
	physical response errors obtained from the final recommendations. Across both
	simulation budgets, MUTACO produces substantially smaller response errors than
	the SBI methods, consistent with its direct allocation of simulator calls
	toward reproducing the prescribed macroscopic response rather than toward
	learning a posterior distribution over parameters.

	At $B=40$, the mean held-out response errors are $2.23\times 10^{-2}$ for MUTACO,
	$1.32\times 10^{-1}$ for NPE, $1.39\times 10^{-1}$ for SNPE, $9.61\times 10^{-2}$ for NLE, and $2.354\times 10^{-1}$ for
	NRE. At $B=100$, the corresponding means are $2.27\times 10^{-2}$, $6.84\times 10^{-2}$, $7.38\times 10^{-2}$,
	$6.75\times 10^{-2}$, and $1.62\times 10^{-1}$. The SBI methods generally improve as the available
	simulation budget increases, particularly NPE, SNPE, and NLE, but the response
	error attained by MUTACO remains substantially lower at both budgets.

	The NPE and SNPE results and the NLE and NRE results originate from separately
	validated SBI benchmark campaigns but are evaluated on the same  target
	cohort and common response metric; the comparison is therefore presented as a
	shared-assessment benchmark rather than as a single jointly executed five-arm experiment.
	The MUTACO entries in the SBI comparison come from a separate MUTACO campaign on the same 120-target cohort and use separate terminal recommendations and held-out assessments; their aggregate values therefore need not match those in the primary comparison.

	\begin{figure*}[t]
		\centering
		\includegraphics[width=0.82\textwidth]
		{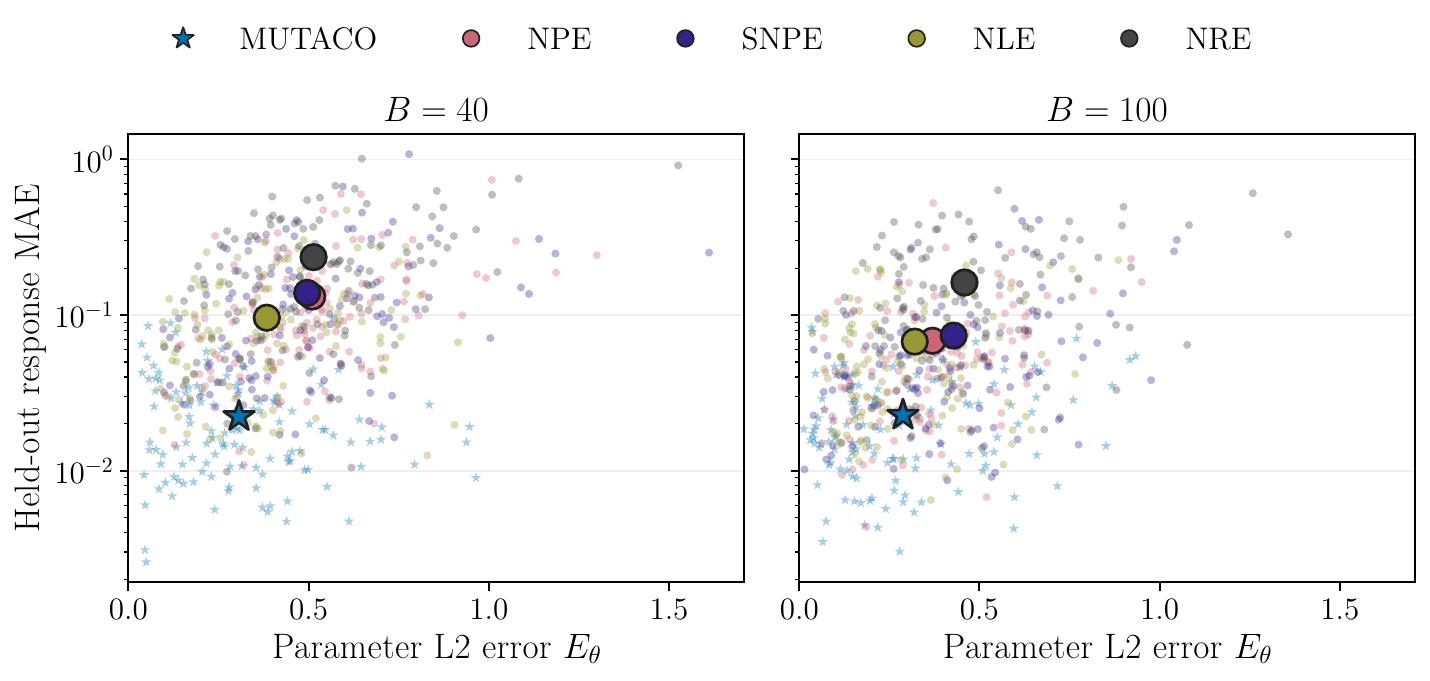}
		\caption{
			Joint comparison of held-out response fidelity and parameter recovery for
			MUTACO, NPE, SNPE, NLE, and NRE at $B=40$ and $B=100$.
			The vertical axis shows the held-out physical response MAE, while the
			horizontal axis shows the Euclidean parameter-recovery error
			$\|\widehat{\boldsymbol{\theta}}-\boldsymbol{\theta}^{\star}\|_2$.
			Each small marker represents one target instance, and the larger denotes the arithmetic mean for the corresponding method in each metric. The comparison illustrates that proximity to the generating
			parameter vector and fidelity to the target macroscopic response are
			related but distinct measures of inverse-problem performance.
		}
		\label{fig:sbi_response_parameter}
	\end{figure*}

	\subsection{Response fidelity and parameter identifiability}
	\label{sec:response_parameter_relation}

	The preceding comparisons show that response reconstruction and recovery of
	the generating disorder parameters are not equivalent measures of inverse
	performance, consistent with the broader distinction between predictive
	fidelity and parameter identifiability in inverse problems
	~\cite{raue_structural_identifiability}.
	Fig.~\ref{fig:sbi_response_parameter} makes this distinction explicit by
	displaying the held-out physical response error together with the distance
	from the generating parameter vector for MUTACO and the four SBI methods.

	The distinction is already apparent at $B=40$. MUTACO has a mean parameter
	error of $3.07\times 10^{-1}$ and a mean held-out response error of $2.23\times 10^{-2}$, whereas NLE,
	the closest SBI method in parameter space, has corresponding mean errors of
	$3.84\times 10^{-1}$ and $9.61\times 10^{-2}$. At $B=100$, the mean parameter errors of MUTACO and NLE
	become relatively close, $2.88\times 10^{-1}$ and $3.20\times 10^{-1}$, respectively, while their mean
	response errors remain substantially separated, $2.27\times 10^{-2}$ and $6.75\times 10^{-2}$.
	Thus, improved recovery of the generating parameters does not necessarily
	produce a proportionate improvement in the macroscopic response.

	Conversely, MUTACO can reproduce the prescribed collective behavior accurately
	without requiring exact recovery of
	$\boldsymbol{\theta}^{\star}$.
	This suggests that the forward response map contains directions in parameter space along which the measured collective response changes only weakly, consistent with the weakly sensitive parameter combinations encountered in sloppy models.~\cite{gutenkunst_sloppiness}. The inverse problem therefore contains an
	identifiability component that cannot be inferred from response error alone.
	To test this directly, we evaluated the local response
	sensitivity at a deterministic subset of 12 well-separated (maximin) targets selected
	from the predefined 120-target benchmark cohort using a maximin criterion. The targets were selected using only their normalized
	generating parameter vectors, without reference to MUTACO performance,
	response error, parameter error, SBI results, or sensitivity information.

\begin{figure}[t]
		\centering
		\includegraphics[width=\columnwidth]{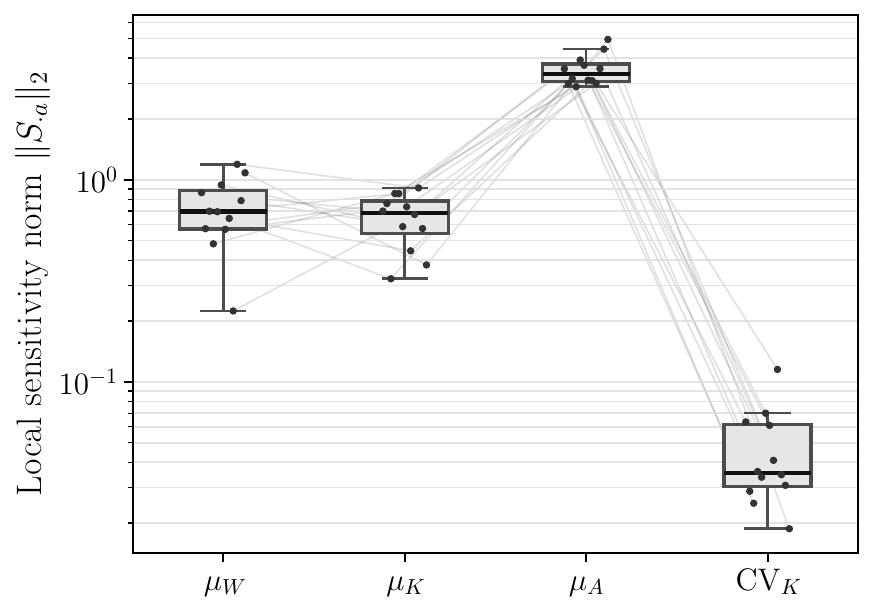}
		\caption{
			Local sensitivity of the collective response across 12 well-separated
			(maximin) targets selected from the predefined 120-target benchmark cohort.
			Dimensionless sensitivity norms $\|S_{\cdot a}\|_2$ are shown for the four
			inferred disorder parameters; individual points denote targets. The target
			subset was selected solely from the normalized generating parameter vectors,
			and boundary-near coordinates use deterministically reduced symmetric
			finite-difference steps.
		}
		\label{fig:local_sensitivity}
	\end{figure}

	For each selected target, we define the dimensionless local sensitivity matrix
	\begin{equation}
		S_{p a}
		=
		\frac{h_a-\ell_a}{s_p}
		\frac{\partial y_p}{\partial \theta_a},
		\qquad
		p=1,\ldots,d_y,
		\quad
		a=1,\ldots,d_\theta ,
		\label{eq:local_sensitivity}
	\end{equation}
	where $h_a-\ell_a$ is the admissible range of parameter coordinate $a$ and
	$s_p$ is the fixed physical scale associated with response component
	$p$. The derivatives are estimated by symmetric finite differences using
	common random numbers within each target to reduce stochastic sampling noise.
	Fig.~\ref{fig:local_sensitivity} summarizes the resulting local response
	sensitivity.

	A consistent sensitivity hierarchy appears across the entire subset.
	The mean bias scale $\mu_A$ is the most sensitive parameter for all 12
	targets, whereas the coupling coefficient of variation $\mathrm{CV}_K$ is
	the least sensitive for all 12. The median sensitivity norms are
	$3.35$ for $\mu_A$, $6.97\times 10^{-1}$ for $\mu_W$, $6.86\times 10^{-1}$ for $\mu_K$, and only
	$3.53\times 10^{-2}$ for $\mathrm{CV}_K$. Moreover, $\mathrm{CV}_K$ provides the
	largest component of the weakest right-singular direction at every target,
	showing that variations in the coupling heterogeneity are systematically the
	least constrained by the collective response. The illustrative target 98
	exhibits the same sensitivity hierarchy, indicating that its local behavior
	is not exceptional within the sampled parameter domain.

	The weak $CV_K$ direction provides a concrete explanation for part of the
	response--parameter mismatch observed above. The chosen coarse observables,
	namely the mean rotation and time-averaged Kuramoto order-parameter magnitude
	over the measured bias conditions, carry substantially less local information
	about the width of the coupling distribution than about the mean bias scale.
	The local geometry also contains parameter-compensation effects: for example,
	the response-sensitivity columns associated with $\mu_W$ and $\mu_A$ are often
	anticorrelated, with cosine similarity reaching approximately $-0.96$ within
	the selected subset. Thus, the mismatch between response fidelity and
	parameter recovery is not solely an optimization effect; it is also a
	property of the forward response map. These results identify locally weakly sensitive directions in the response map over the sampled parameter domain, without establishing global structural non-identifiability~\cite{chis_sloppiness_identifiability}.

	For the present application, this distinction is physically important.
	If the objective is to reproduce or control a macroscopic response, distance
	from the particular parameter vector used to generate the synthetic target is
	not by itself a sufficient measure of success. MUTACO is designed around this
	response-focused objective, while the SBI methods place greater emphasis on
	recovering the parameter distribution implied by the observations. The two
	tasks are complementary, but Fig.~\ref{fig:sbi_response_parameter} shows that
	they need not rank candidate solutions in the same way.

	\section{Discussion}
	\label{sec:discussion}

	The results show that response-focused inversion and parameter identification
	should be distinguished in stochastic oscillator networks. Across the
	benchmark ensemble, MUTACO consistently reproduces the prescribed macroscopic
	responses with lower held-out error than both objective-aligned Bayesian
	optimization and simulation-based inference methods, while also recovering
	the generating disorder parameters comparatively well. At the same time, the
	joint response -- parameter analysis shows that small parameter error and small
	response error are not interchangeable: distinct ensemble-parameter
	configurations can generate closely similar collective behavior over the
	measured bias conditions.

    {The methods differ in several coupled respects. MUTACO retains the
multicomponent response and uses learned correlations and posterior uncertainty
to organize sequential search.} Scalar qLogNEI instead compresses each simulator evaluation to a
single target-error value before fitting its surrogate. That scalarization is
objective aligned, but it discards information about how the individual
response components approach the target and how their predictive uncertainties
are correlated. MUTACO uses this richer surrogate to define an
uncertainty-dependent compatibility region, excluding presently incompatible
candidates from target-directed refinement while allowing larger apparent
mismatches where the response remains uncertain. Within the compatible region,
a logarithmic barrier guides refinement away from its boundary, whereas failure
to identify a compatible candidate triggers exploration toward the region of
largest posterior response uncertainty. This distinction is particularly
relevant under severe simulator budgets, consistent with the larger performance
gap observed at $B=40$ than at $B=100$. {This comparison evaluates the
complete procedures rather than isolating individual acquisition components.}

	The local-sensitivity analysis shows that these observables constrain different
	parameter directions very unevenly. Consequently, accurate reconstruction of
	the measured response can coexist with substantial uncertainty along locally
	weak parameter directions, as occurs more generally in sloppy parameterized
	models~\cite{monsalve_bravo_sloppiness}. This response--parameter mismatch
	clarifies the role of identifiability in the present problem: the target vector
	contains only coarse rotation and synchronization observables measured at a
	finite set of bias conditions, and parameter vectors separated in the
	four-dimensional inference space can therefore remain close in response space.
	From the perspective of response reproduction or control, such solutions are
	not necessarily failures if they generate nearly indistinguishable macroscopic
	behavior over the operating conditions of interest. Parameter recovery and
	response fidelity should therefore be reported separately rather than collapsed
	into a single notion of inverse accuracy.

	The comparison with simulation-based inference emphasizes the same
	distinction from another direction. SBI methods are designed to characterize
	parameter distributions conditioned on observations, whereas MUTACO allocates
	its limited simulator budget directly toward locating parameter vectors that
	reproduce a specified response. The substantially lower response errors of
	MUTACO therefore should not be interpreted as a general statement that
	response-focused optimization is preferable to posterior inference. Rather,
	the appropriate formulation depends on the scientific objective. Posterior
	inference is essential when uncertainty in the underlying parameters is itself
	the quantity of interest; direct response matching is attractive when the
	primary goal is to reproduce, design, or control an observable collective
	behavior under a restricted simulation budget.

The terminal-domain ablation in Appendix~\ref{app:terminal} further separates
the roles of acquisition and final recommendation. Holding the completed
$B=40$ MUTACO trajectories and fitted surrogates fixed, continuous
optimization of the final posterior-mean response error reduces the mean
held-out terminal error by approximately $28.5\%$ relative to selecting only
among previously evaluated points, without requiring additional simulator
evaluations. The improvement therefore reflects additional exploitation of the
completed surrogate rather than a change in the sequential acquisition
trajectory. Appendix~\ref{app:terminal} gives the corresponding paired
comparison.

The absence of derivative requirements broadens the class of forward models
that can be treated in this way. MUTACO requires only a queryable stochastic
simulator returning a multicomponent response; it does not require an
analytically differentiable forward model, accessible response Jacobians,
adjoint construction, or governing equations exposed to the optimizer. This
does not imply that the response map may be arbitrarily irregular: successful
surrogate-based search still requires that its relevant structure be
learnable from the available simulation budget.

	The present formulation already accommodates stochastic variability in the
	forward response; changes in network size or effective noise intensity, as well as additional
	measurement noise, primarily modify the response statistics rather than the
	structure of the inverse problem. More substantive extensions concern regimes
	with strongly non-Gaussian response statistics or chaotic dynamics, where the
	present surrogate description may become inadequate. Scaling to substantially
	higher-dimensional inverse problems is another open direction and will likely
	require an appropriate low-dimensional encoding or identification of the
	dominant parameter directions. Robustness to model mismatch remains the usual
	separate question associated with any model-based inverse problem.

	More broadly, the relevant requirement is that the parameter-to-response map
	contain structure that can be learned from the available simulation budget.
	When the scientific objective is faithful reproduction of collective behavior
	rather than unique microscopic identification, sequential design can then
	benefit from preserving the multicomponent response and its predictive
	uncertainty rather than reducing each simulator evaluation immediately to a
	scalar mismatch.

	\section{Conclusion}
	\label{sec:conclusion}

	We introduced MUTACO, a compatibility-aware sequential-design method for
	response-focused inversion with stochastic simulators. Applied to a heterogeneous stochastic oscillator network, MUTACO consistently achieved lower held-out response
	errors than objective-aligned Bayesian optimization, nonadaptive Sobol search,
	and several simulation-based inference methods under fixed simulator budgets.
	It also recovered the generating disorder parameters comparatively well,
	while the joint response--parameter analysis showed that accurate reproduction
	of collective behavior does not require unique recovery of the underlying
	ensemble parameters.

	These results support a response-centered view of inverse modeling in
	stochastic dynamical systems. A central feature of MUTACO is that it requires
	neither response gradients nor adjoint sensitivities and does not require the
	simulator to expose a differentiable model or its governing equations. The
	method operates directly from finite noisy evaluations of a multicomponent
	stochastic response, making it applicable to simulator-mediated inverse
	problems in which reproducing observable collective behavior is more relevant
	than uniquely identifying the microscopic parameters that generated it.

	\begin{acknowledgments}
		A.C.K. is deeply grateful to
		Dan MacKinlay and Laurence Davies for introducing him to simulation-based
		inference for inverse problems and for their generous time and many discussions.
		These interactions, all predating the present work, helped shape his thinking
		about inverse problems and their computational treatment.
		This publication has been produced with support from the TÜBİTAK 2232/A International Fellowship for Outstanding Researchers Program (Project No.~124C530); A.C.K. and U.C.T. acknowledge support under this project. The work of İ.O.K. and C.T. was supported in part by TÜBİTAK Grant No.~124E065. İ.O.K. was also supported by Türk Telekom as part of the 5G and Beyond Joint Graduate Support Programme coordinated by the Information and Communication Technologies Authority. C.T. acknowledges support by the Turkish Academy of Sciences Distinguished Young Scientist Award Program (TÜBA-GEBİP-2023) and the TÜBİTAK 2024 Incentive Award. The numerical calculations reported in this paper were performed at TÜBİTAK ULAKBİM, High Performance and Grid Computing Center (TRUBA resources). All responsibility for the content of the publication rests with the authors and is not scientifically endorsed by TÜBİTAK.

		OpenAI's ChatGPT and Codex were used to assist with software development and debugging, computational workflow preparation and scheduling on TRUBA, bibliographic verification, mathematical consistency checks, and manuscript preparation. The tools were employed through sustained, real-time, iterative exchanges under the authors' direction, involving critical examination of intermediate outputs, challenges to proposed solutions and assumptions, and successive refinements. AI-assisted outputs were subjected to adversarial auditing for errors, inconsistencies, and unsupported claims. The authors independently and manually verified the relevant outputs against computational tests, archived results, and primary literature, as appropriate. The scientific formulation, methodological decisions, interpretation of results, and final responsibility for the work remain with the authors.
	\end{acknowledgments}

	\section*{Code and Data Availability}
	
	The source code and data supporting the results
	of this work will be made publicly available at
	\url{https://github.com/aydinCkeser/MUTACO}.

	\appendix

	\section{Gaussian-Process Theory Background}
	\label{app:gp_theory}

	For completeness, we summarize the Gaussian-process conditioning framework
	underlying the surrogate construction used in this work and its connection to
	Gaussian field theory. We use a scalar response field $f(x)$ for notational
	clarity; the extension to the vector-valued multitask process used by MUTACO is
	obtained by promoting the covariance kernel to a matrix-valued kernel, as
	described in Appendix~\ref{app:gp}. Standard treatments of Gaussian processes
	and Gaussian fields may be found in
	Refs.~\cite{rasmussen_williams_gp,kardar_statistical_fields}.

	We use both operator inner-product notation and, where convenient, a compact
	index notation. Specifically,
	\begin{equation}
		\langle g,\mathcal A f\rangle
		\equiv
		\int dx\,dx'\,
		g(x)\mathcal A(x,x')f(x'),
	\end{equation}
	while in condensed notation the same expression is written
	$g^a\mathcal A_{ab}f^b$, with repeated continuous indices implying
	integration over the corresponding coordinates.

	A Gaussian process assigns a probability measure to a space of functions.
	Formally, expectation values of a functional $\mathcal O[f]$ may be written as
	\begin{equation}
		\left\langle \mathcal O[f] \right\rangle
		=
		\frac{1}{Z_f}
		\int \mathcal Df\,
		\mathcal O[f]\,
		e^{-S_f[f]},
		\label{eq:app_gp_functional_average}
	\end{equation}
	where $Z_f$ normalizes the measure. For a Gaussian process with mean
	$m(x)$ and covariance kernel $k(x,x')$, the action is quadratic,
	$S_f[f]=\frac{1}{2}\langle f-m,k^{-1}(f-m)\rangle$, where the inner
	product denotes integration over the input domain. Equivalently,
	\begin{equation}
		P[f]
		=
		\frac{1}{Z_f}
		\exp\!\left[
		-\frac{1}{2}
		\left\langle
		f-m,\,
		k^{-1}(f-m)
		\right\rangle
		\right].
		\label{eq:app_gp_prior}
	\end{equation}
	For a real-valued process the covariance is symmetric,
	$k(x,x')=k(x',x)$, and its inverse is defined by
	$\int dx''\,k^{-1}(x,x'')k(x'',x')=\delta(x-x')$.

	In this representation, $k$ is the two-point correlation function of the
	field,
	\begin{equation}
		\left\langle
		[f(x)-m(x)]
		[f(x')-m(x')]
		\right\rangle
		=
		k(x,x'),
		\label{eq:app_gp_twopoint}
	\end{equation}
	and $k^{-1}$ is the corresponding quadratic, or precision, operator. This is
	the same mathematical structure as a free Gaussian field theory: $k$ plays the
	role of a propagator and $k^{-1}$ the role of an inverse propagator. The analogy
	is statistical rather than dynamical; the coordinate $x$ need not represent
	physical position or time.

	Suppose now that the field is observed at locations
	$X=\{x_1,\ldots,x_n\}$. The measurements satisfy
	\begin{equation}
		y_i
		=
		f(x_i)+\epsilon_i,
		\label{eq:app_gp_observation}
	\end{equation}
	where the observation noise is Gaussian,
	$\boldsymbol{\epsilon}
	\sim
	\mathcal N(\mathbf 0,\boldsymbol{\Sigma}_{\epsilon})$. Its quadratic action is
	\begin{equation}
		S_{\epsilon}
		=
		\frac{1}{2}
		\boldsymbol{\epsilon}^{\mathsf T}
		\boldsymbol{\Sigma}_{\epsilon}^{-1}
		\boldsymbol{\epsilon},
		\label{eq:app_gp_noise_action}
	\end{equation}
	and hence the joint prior measure is
	\begin{equation}
		P[f,\boldsymbol{\epsilon}]
		\propto
		\exp\!\left[
		-S_f[f]-S_{\epsilon}[\boldsymbol{\epsilon}]
		\right].
		\label{eq:app_gp_joint_measure}
	\end{equation}

	It is useful to introduce a restriction operator $R$ that evaluates the field
	at the observation points,
	\begin{equation}
		\mathbf f_X = Rf,
		\qquad
		(\mathbf f_X)_i=f(x_i).
	\end{equation}
	The observation equation is then
	$\mathbf y=Rf+\boldsymbol{\epsilon}$. Eliminating the unobserved noise gives
	the Gaussian likelihood
	\begin{equation}
		P(\mathbf y\mid f)
		\propto
		\exp\!\left[-S_{\rm obs}[f;\mathbf y]\right],
	\end{equation}
	with
	\begin{equation}
		S_{\rm obs}[f;\mathbf y]
		=
		\frac{1}{2}
		(\mathbf y-Rf)^{\mathsf T}
		\boldsymbol{\Sigma}_{\epsilon}^{-1}
		(\mathbf y-Rf).
		\label{eq:app_gp_observation_action}
	\end{equation}

	Bayes' rule therefore gives the posterior functional density
	\begin{equation}
		P[f\mid\mathbf y]
		=
		\frac{1}{\mathcal Z_{\mathbf y}}
		\exp\!\left[
		-S_{\rm post}[f;\mathbf y]
		\right],
		\qquad
		S_{\rm post}
		=
		S_f+S_{\rm obs},
		\label{eq:app_gp_posterior_action}
	\end{equation}
	where $\mathcal Z_{\mathbf y}$ is the posterior normalization. Because both
	terms in $S_{\rm post}$ are quadratic in $f$, conditioning preserves the
	Gaussian form.

	Collecting terms according to their dependence on the field gives
	\begin{equation}
		S_{\rm post}[f;\mathbf y]
		=
		\frac{1}{2}
		\langle f,\mathcal A f\rangle
		-
		\langle J_{\mathbf y},f\rangle
		+
		C(\mathbf y),
		\label{eq:app_gp_quadratic_posterior}
	\end{equation}
	where
	\begin{equation}
		\mathcal A
		=
		k^{-1}
		+
		R^{\mathsf T}\boldsymbol{\Sigma}_{\epsilon}^{-1}R,
		\label{eq:app_gp_precision}
	\end{equation}
	\begin{equation}
		J_{\mathbf y}
		=
		k^{-1}m
		+
		R^{\mathsf T}\boldsymbol{\Sigma}_{\epsilon}^{-1}\mathbf y,
		\label{eq:app_gp_source}
	\end{equation}
	and
	\begin{equation}
		C(\mathbf y)
		=
		\frac{1}{2}
		\langle m,k^{-1}m\rangle
		+
		\frac{1}{2}
		\mathbf y^{\mathsf T}\boldsymbol{\Sigma}_{\epsilon}^{-1}\mathbf y .
		\label{eq:app_gp_constant}
	\end{equation}
	Thus the observations modify the prior precision operator from $k^{-1}$ to
	$\mathcal A$ and generate the effective source $J_{\mathbf y}$. The
	field-independent term $C(\mathbf y)$ contributes to the normalization but
	does not affect the posterior mean or covariance.

	Let
	\begin{equation}
		\mu(x)
		=
		\left\langle f(x)\right\rangle_{\mathbf y}
	\end{equation}
	denote the posterior mean and
	\begin{equation}
		\kappa(x,x')
		=
		\left\langle
		[f(x)-\mu(x)]
		[f(x')-\mu(x')]
		\right\rangle_{\mathbf y}
	\end{equation}
	the posterior covariance. Since the posterior action is Gaussian, these
	quantities are determined by
	\begin{equation}
		\mathcal A\,\mu
		=
		J_{\mathbf y}
		,
		\qquad
		\mathcal A\,\kappa
		=
		I.
		\label{eq:app_gp_mean_cov_equations}
	\end{equation}
	Equivalently,
	\begin{equation}
		\mu
		=
		\mathcal A^{-1}J_{\mathbf y},
		\qquad
		\kappa
		=
		\mathcal A^{-1}.
		\label{eq:app_gp_mean_cov_operator}
	\end{equation}
	In field-theory language, $\mathcal A$ is the dressed inverse propagator,
	$\kappa$ is the posterior propagator, and $J_{\mathbf y}$ is the
	data-induced source that shifts the center of the Gaussian field.

	For the posterior covariance, one need not invert the continuum operator
	$\mathcal A(x,x')$ directly. The Woodbury identity reduces the calculation
	to inversion of the finite covariance matrix defined on the observed inputs and gives
	\begin{equation}
		\left(
		k^{-1}+R^{\mathsf T}\boldsymbol{\Sigma}_{\epsilon}^{-1}R
		\right)^{-1}
		=
		k
		-
		kR^{\mathsf T}
		\left(
		\boldsymbol{\Sigma}_{\epsilon}+RkR^{\mathsf T}
		\right)^{-1}
		Rk .
		\label{eq:app_gp_woodbury}
	\end{equation}
	Defining the finite covariance matrix and prior-mean vector at the observation
	points by
	\begin{equation}
		K_{ij}=k(x_i,x_j),
		\qquad
		M_i=m(x_i),
	\end{equation}
	and the cross-covariance vectors
	\begin{equation}
		[\mathbf k_{xX}]_i=k(x,x_i),
		\qquad
		\mathbf k_{Xx'}=\mathbf k_{x'X}^{\mathsf T},
	\end{equation}
	one obtains the standard Gaussian-process conditioning relations
	\begin{equation}
		\mu(x)
		=
		m(x)
		+
		\mathbf k_{xX}
		(K+\boldsymbol{\Sigma}_{\epsilon})^{-1}
		(\mathbf y-\mathbf M)
		,
		\label{eq:app_gp_standard_mean}
	\end{equation}
	and
	\begin{equation}
		\kappa(x,x')
		=
		k(x,x')
		-
		\mathbf k_{xX}
		(K+\boldsymbol{\Sigma}_{\epsilon})^{-1}
		\mathbf k_{Xx'}
		.
		\label{eq:app_gp_standard_cov}
	\end{equation}
	The posterior variance at a single input is consequently
	\begin{equation}
		\sigma_{\rm post}^{2}(x)
		=
		\kappa(x,x).
		\label{eq:app_gp_standard_variance}
	\end{equation}
Conditional on fixed kernel and observation-noise hyperparameters,
Eqs.~\eqref{eq:app_gp_standard_mean} and
\eqref{eq:app_gp_standard_cov} show explicitly how conditioning separates into
two effects. The observed values $\mathbf y$ shift the posterior mean,
	whereas the posterior covariance depends on the locations of the observations,
	the prior kernel, and the observation-noise covariance, but not on the
	numerical values of the observations themselves. Further derivations and discussion of Gaussian-process conditioning can be
	found in Ref.~\cite{rasmussen_williams_gp}; the corresponding Gaussian-field
	interpretation follows the standard quadratic-field construction reviewed in
	Ref.~\cite{kardar_statistical_fields}.

	The following appendix specializes these general relations to the correlated
	multitask Gaussian process, response normalization, and acquisition rules used
	in MUTACO.

	\section{MUTACO Gaussian-Process and Acquisition Implementation}
	\label{app:gp}

	\subsection{Multitask Gaussian-process surrogate}
	\label{app:gp_surrogate}

	We now specialize the Gaussian-process conditioning framework of
	Appendix~\ref{app:gp_theory} to the correlated multitask surrogate used in
	the MUTACO calculations
	\cite{rasmussen_williams_gp,alvarez_multitask_gp,bonilla_multitask_gp}.
	Lowercase Latin indices $a,b=1,\ldots,4$ denote parameter-space
coordinates, $p,p'=1,\ldots,d_y$ denote response components, and
$i,j$ denote simulator evaluations.

	The physical search domain defined in
	Eq.~\eqref{eq:inverse_bounds},
	\begin{equation}
		\Theta=[0.5,2]^3\times[0,1]
	\end{equation}
	is mapped affinely to the unit four-cube $[0,1]^4$ before GP fitting. For
	each parameter coordinate,
	\begin{equation}
		u^a
		=
		\frac{\theta^a-\ell^a}{h^a-\ell^a},
		\qquad
		a=1,\ldots,4,
		\label{eq:input_normalization}
	\end{equation}
	with lower- and upper-bound vectors
	\begin{equation}
		\boldsymbol{\ell}=(0.5,0.5,0.5,0),
		\qquad
		\boldsymbol{h}=(2,2,2,1).
		\label{eq:parameter_bounds}
	\end{equation}

	The $d_y=20$ response components are standardized using statistics computed
	from the initial design. For $n_0=16$ initial simulator evaluations, define
	\begin{equation}
		c^p
		=
		\frac{1}{n_0}
		\sum_{i=1}^{n_0} y_i^p ,
		\label{eq:response_center}
	\end{equation}
	and
	\begin{equation}
		s^p
		=
		\max\left[
		\left(
		\frac{1}{n_0-1}
		\sum_{i=1}^{n_0}
		\left(y_i^p-c^p\right)^2
		\right)^{1/2},
		10^{-12}
		\right].
		\label{eq:response_scale}
	\end{equation}
	The observations and target are represented in standardized response
	coordinates as
	\begin{equation}
		\widetilde y_i^p
		=
		\frac{y_i^p-c^p}{s^p},
		\qquad
		\widetilde y_\star^p
		=
		\frac{y_\star^p-c^p}{s^p}.
		\label{eq:response_standardization}
	\end{equation}
	The quantities $c^p$ and $s^p$ are computed once from the initial design and
	held fixed throughout the sequential search, so that the response-space
	geometry does not change as new simulator evaluations are acquired. The
	floor in Eq.~\eqref{eq:response_scale} prevents division by a numerically
	vanishing initial-design scale.

	Let $f^p(\mathbf u)$ denote latent standardized response component $p$.
	For the calculations reported here, the multitask GP has a task-dependent
	constant mean,
	\begin{equation}
		\mathbb E[f^p(\mathbf u)]
		=
		m^p,
	\end{equation}
	and a separable covariance
	\begin{equation}
		k^{pp'}(\mathbf u,\mathbf u')
		=
		k_{\mathrm{RBF,ARD}}(\mathbf u,\mathbf u')\,B^{pp'},
		\label{eq:multitask_kernel}
	\end{equation}
	where
	\begin{equation}
		k_{\mathrm{RBF,ARD}}(\mathbf u,\mathbf u')
		=
		\exp\!\left[
		-\frac{1}{2}
		\sum_{a=1}^{4}
		\frac{(u^a-u'^a)^2}{\lambda_a^2}
		\right].
		\label{eq:ard_kernel}
	\end{equation}
	Here $\lambda_a$ is the characteristic length scale along normalized
	parameter direction $a$~\cite{rasmussen_williams_gp}. Because all four
	coordinates have been mapped to $[0,1]$, these length scales characterize
	response variation relative to a common parameter-space scale: small
	$\lambda_a$ permits rapid variation along coordinate $a$, whereas large
	$\lambda_a$ corresponds to a slowly varying response. The matrix $B^{pp'}$ is
	a learned positive-semidefinite task-covariance matrix coupling the response
	components~\cite{alvarez_multitask_gp,bonilla_multitask_gp}. For the benchmark calculations, $B^{pp'}$ is parameterized as a rank-one
	contribution plus a diagonal component.

	For training inputs $\{\mathbf u_i\}_{i=1}^{n}$, the prior covariance between
	response component $p$ at $\mathbf u_i$ and component $p'$ at $\mathbf u_j$ is
	\begin{equation}
		K^{pp'}_{ij}
		=
		k^{pp'}(\mathbf u_i,\mathbf u_j).
		\label{eq:multioutput_training_covariance}
	\end{equation}
	The standardized simulator responses are modeled as
	\begin{equation}
		\widetilde y_i^p
		=
		f^p(\mathbf u_i)+\epsilon_i^p ,
	\end{equation}
	where the observation noise is taken to be zero-mean Gaussian with covariance
	\begin{equation}
		\left\langle
		\epsilon_i^p\epsilon_j^{p'}
		\right\rangle
		=
		\left[\boldsymbol{\Sigma}_{\epsilon}\right]^{pp'}_{ij}
		=
		\delta_{ij}\delta^{pp'}\sigma_p^2 .
		\label{eq:gp_noise_covariance}
	\end{equation}
	A separate observation-noise variance $\sigma_p^2$ is therefore learned for
	each response component, while cross-component observation-noise correlations
	are not included. This noise model accounts for the finite-sample variability
	of the stochastic simulator responses within the GP likelihood. The covariance
	of the observed training data is consequently
	\begin{equation}
		C^{pp'}_{ij}
		=
		K^{pp'}_{ij}
		+
		\left[\boldsymbol{\Sigma}_{\epsilon}\right]^{pp'}_{ij}.
		\label{eq:gp_observed_covariance}
	\end{equation}

	The pair $(i,p)$ may be treated as a single compound index, so that
	$C^{pp'}_{ij}$ is represented by an $nd_y\times nd_y$ covariance matrix
	$\mathbf C$. We likewise stack the standardized observations into the
	$nd_y$-dimensional vector $\widetilde{\mathbf y}$ and the corresponding prior
	means into $\mathbf M$. The training-data distribution is then
	\begin{equation}
		\widetilde{\mathbf y}
		\sim
		\mathcal N
		\left(
		\mathbf M,\mathbf C
		\right).
		\label{eq:gp_training_distribution}
	\end{equation}

	For a candidate point $\mathbf u$, let
	$\mathbf K_{\mathbf u X}$ denote the covariance between its $d_y$ latent
	response components and the $nd_y$ latent training values, with elements
	\begin{equation}
		\left[
		\mathbf K_{\mathbf u X}
		\right]_{p,jp'}
		=
		k^{pp'}(\mathbf u,\mathbf u_j).
		\label{eq:gp_cross_covariance}
	\end{equation}
	The corresponding $d_y\times d_y$ prior covariance at the candidate point is
	\begin{equation}
		\left[
		\mathbf K_{\mathbf u\mathbf u}
		\right]_{pp'}
		=
		k^{pp'}(\mathbf u,\mathbf u).
		\label{eq:gp_candidate_covariance}
	\end{equation}
	By covariance symmetry,
	$\mathbf K_{X\mathbf u}=\mathbf K_{\mathbf u X}^{T}$.

	Conditioning the joint Gaussian distribution on the observed training data
	gives the latent posterior mean
	\begin{equation}
		\widetilde{\boldsymbol{\mu}}_n(\mathbf u)
		=
		\mathbf m(\mathbf u)
		+
		\mathbf K_{\mathbf u X}
		\mathbf C^{-1}
		\left(
		\widetilde{\mathbf y}-\mathbf M
		\right),
		\label{eq:gp_posterior_mean}
	\end{equation}
	and latent posterior covariance
	\begin{equation}
		\widetilde{\boldsymbol{\Sigma}}_n(\mathbf u)
		=
		\mathbf K_{\mathbf u\mathbf u}
		-
		\mathbf K_{\mathbf u X}
		\mathbf C^{-1}
		\mathbf K_{X\mathbf u}.
		\label{eq:gp_posterior_covariance}
	\end{equation}
	Here
	$\widetilde{\boldsymbol{\mu}}_n(\mathbf u)\in\mathbb R^{d_y}$ and
	$\widetilde{\boldsymbol{\Sigma}}_n(\mathbf u)\in
	\mathbb R^{d_y\times d_y}$ describe the latent multicomponent response at
	one candidate point. In particular,
	$\widetilde{\boldsymbol{\Sigma}}_n$ excludes the learned observation-noise
	contribution and is the uncertainty entering the MUTACO compatibility and
	exploration rules below. The inverse notation in
	Eqs.~\eqref{eq:gp_posterior_mean} and \eqref{eq:gp_posterior_covariance}
	denotes the corresponding covariance-system solves; the matrix inverse is
	not formed explicitly in the numerical implementation. Numerical
	factorization, precision handling, and recovery procedures are described in
	Appendix~\ref{app:numerics}.

	The GP hyperparameters---including the four parameter-space length scales,
	the task covariance, task means, and task-specific observation-noise
	variances---are learned by maximizing the exact marginal likelihood
	~\cite{rasmussen_williams_gp}. Writing
	$\boldsymbol{\vartheta}$ for these hyperparameters, the log marginal
	likelihood is
	\begin{align}
		\log p(
		\widetilde{\mathbf y}
		\mid\boldsymbol{\vartheta}
		)
		={}&
		-\frac{1}{2}
		\left(
		\widetilde{\mathbf y}-\mathbf M
		\right)^T
		\mathbf C^{-1}
		\left(
		\widetilde{\mathbf y}-\mathbf M
		\right)
		\nonumber\\
		&-\frac{1}{2}\log\det\mathbf C
		-\frac{nd_y}{2}\log(2\pi),
		\label{eq:gp_marginal_likelihood}
	\end{align}
	where $\mathbf M$ and $\mathbf C$ depend on
	$\boldsymbol{\vartheta}$. The surrogate is refit after each new simulator
	evaluation. Numerical details of the marginal-likelihood optimization,
	precision handling, initialization, and bounded recovery are given in
	Appendix~\ref{app:numerics}.

	\subsection{Target compatibility and sequential acquisition}
	\label{app:gp_acquisition}

	At sequential step $t$, let
	$\widetilde{\boldsymbol{\mu}}_t(\boldsymbol{\theta})$ and
	$\widetilde{\boldsymbol{\Sigma}}_t(\boldsymbol{\theta})$ denote the latent
	posterior mean and covariance evaluated at the normalized coordinate
	$\mathbf u(\boldsymbol{\theta})$ corresponding to the physical parameter
	vector $\boldsymbol{\theta}$.

	The residual between the posterior mean and the standardized target is
	\begin{equation}
		\mathbf r_t(\boldsymbol{\theta})
		=
		\widetilde{\boldsymbol{\mu}}_t(\boldsymbol{\theta})
		-
		\widetilde{\mathbf y}^{\star},
		\label{eq:mutaco_residual}
	\end{equation}
	or, componentwise,
	\begin{equation}
		r_t^p(\boldsymbol{\theta})
		=
		\widetilde{\mu}_t^p(\boldsymbol{\theta})
		-
		\widetilde y_\star^p .
	\end{equation}
	Its squared Euclidean magnitude is
	\begin{equation}
		q_t(\boldsymbol{\theta})
		=
		\sum_{p=1}^{d_y}
		\left[
		r_t^p(\boldsymbol{\theta})
		\right]^2 .
		\label{eq:mutaco_residual_norm}
	\end{equation}

	To characterize posterior uncertainty in the full response space, MUTACO
	uses the largest eigenvalue of the latent response covariance,
	\begin{equation}
		\lambda_t(\boldsymbol{\theta})
		=
		\lambda_{\max}
		\left[
		\widetilde{\boldsymbol{\Sigma}}_t(\boldsymbol{\theta})
		\right].
		\label{eq:mutaco_lambda}
	\end{equation}
	Because
	$\widetilde{\boldsymbol{\Sigma}}_t\in\mathbb R^{d_y\times d_y}$,
	$\lambda_t$ measures the largest posterior variance over all directions in
	the correlated response space.

	{The} uncertainty multiplier is set from the response dimension according to
	\begin{equation}
		\beta
		=
		{\chi_{d_y}(0.95)}.
		\label{eq:mutaco_beta}
	\end{equation}
	where $\chi_{d_y}(0.95)$ denotes the 95th percentile of the chi distribution with $d_y$ degrees of freedom. This reference quantile sets the scale of the operational compatibility
	criterion; it does not assign a calibrated posterior probability that a
	candidate generates the prescribed target. Consequently, $\beta$ is the radius containing $95\%$ of the probability mass of an isotropic unit-variance Gaussian residual in response space. The resulting threshold therefore accounts explicitly for the number of response components rather than introducing an independently tuned confidence parameter.

	For the present $d_y=20$ response,
	\begin{equation}
		\beta
		\approx
		5.60.
	\end{equation}
	We then define the squared compatibility radius
	\begin{equation}
		\rho_t^2(\boldsymbol{\theta})
		=
		\beta^2
		\lambda_t(\boldsymbol{\theta}).
		\label{eq:mutaco_radius}
	\end{equation}
	A candidate is regarded as compatible with the target whenever
	\begin{equation}
		q_t(\boldsymbol{\theta})
		\leq
		\rho_t^2(\boldsymbol{\theta}),
		\label{eq:mutaco_compatibility}
	\end{equation}
	or equivalently,
	\begin{equation}
		\left\|
		\widetilde{\boldsymbol{\mu}}_t(\boldsymbol{\theta})
		-
		\widetilde{\mathbf y}^{\star}
		\right\|_2^2
		\leq
		\beta^2
		\lambda_{\max}
		\left[
		\widetilde{\boldsymbol{\Sigma}}_t(\boldsymbol{\theta})
		\right].
	\end{equation}
	Equation~\eqref{eq:mutaco_compatibility} defines a spherical compatibility
	bound obtained by enclosing the corresponding chi-square-calibrated Gaussian
	ellipsoid using the largest eigenvalue of the posterior covariance. It is
	therefore a conservative operator-norm bound and should not be confused with
	the full Gaussian credible ellipsoid, which would involve the inverse
	covariance.

	For $\rho_t^2(\boldsymbol{\theta})>0$, it is convenient to introduce the
	normalized compatibility margin
	\begin{equation}
		\Delta_t(\boldsymbol{\theta})
		=
		\frac{
			\rho_t^2(\boldsymbol{\theta})
			-
			q_t(\boldsymbol{\theta})
		}{
			\rho_t^2(\boldsymbol{\theta})
		}
		=
		1-
		\frac{
			q_t(\boldsymbol{\theta})
		}{
			\beta^2\lambda_t(\boldsymbol{\theta})
		}.
		\label{eq:mutaco_margin}
	\end{equation}
	The compatibility condition itself includes the boundary
	$\Delta_t=0$. For numerical refinement, however, the logarithmic barrier
	requires a strict interior. The production implementation therefore admits
	barrier candidates only when
	\begin{equation}
		\Delta_t(\boldsymbol{\theta})>10^{-8}.
		\label{eq:mutaco_barrier_interior}
	\end{equation}

	Within this strict interior, candidate refinement minimizes
	\begin{equation}
		J_t(\boldsymbol{\theta})
		=
		\left\|
		\mathbf r_t(\boldsymbol{\theta})
		\right\|_2
		-
		\alpha
		\log
		\Delta_t(\boldsymbol{\theta}),
		\qquad
		\alpha=0.1.
		\label{eq:mutaco_barrier_objective}
	\end{equation}
	The first term favors candidates whose posterior mean approaches the target,
	whereas the logarithmic term prevents the refinement from approaching the
	compatibility boundary too closely.

	At each sequential step, a scrambled-Sobol screening set of 64 candidate
	points is generated over the admissible parameter domain. If this set
	contains at least one finite candidate satisfying
	Eq.~\eqref{eq:mutaco_barrier_interior}, the candidates are ranked by
	Eq.~\eqref{eq:mutaco_barrier_objective}. Up to the four best screening
	points are then used to initialize bounded continuous refinement. The
	production refinement uses projected gradient descent with feasibility
	backtracking, retaining the best finite non-worsening result from each
	restart.

	If no finite strict-interior candidate is found in the screening set,
	MUTACO switches to uncertainty-driven exploration. The exploratory objective
	is
	\begin{equation}
		U_t(\boldsymbol{\theta})
		=
		\lambda_{\max}
		\left[
		\widetilde{\boldsymbol{\Sigma}}_t(\boldsymbol{\theta})
		\right],
		\label{eq:mutaco_exploration}
	\end{equation}
	and the next candidate is selected approximately according to
	\begin{equation}
		\boldsymbol{\theta}_{t+1}
		\approx
		\arg\max_{\boldsymbol{\theta}\in\Theta}
		U_t(\boldsymbol{\theta}).
	\end{equation}
	The four highest-ranked screening points initialize projected Adam
	refinement of Eq.~\eqref{eq:mutaco_exploration}. Thus, both acquisition
	branches combine a finite Sobol screening stage with bounded continuous
	local refinement.

	The exploration branch is triggered by the absence of a strict-interior
	candidate in the finite screening set. It should therefore not be
	interpreted as establishing that the continuous parameter domain contains no
	compatible point.  Rather, it provides a fallback that directs the next simulator evaluation
	toward regions of greatest posterior uncertainty when the current numerical
	search has not identified a target-compatible candidate.

	After refinement, the selected parameter vector is evaluated with the
	stochastic simulator, the resulting multicomponent response is added to the
	training set, and the multitask GP is refit. If the selected candidate
	coincides exactly with a previously evaluated point, the implementation
	returns the highest-ranked finite unevaluated point from the screening set.
	The procedure is repeated until the prescribed simulator budget is
	exhausted.

	\subsection{Final recommendation}
	\label{app:gp_terminal}

	The terminal recommendation is treated separately from the sequential
	acquisition rule. Once the simulation budget $B$ has been exhausted, the
	multitask GP is fit to the complete dataset and its latent posterior mean is
	converted back from standardized to physical response coordinates according
	to
	\begin{equation}
		\mu^{p,\mathrm{phys}}_B(\boldsymbol{\theta})
		=
		c^p
		+
		s^p\,
		\widetilde{\mu}^{p}_B(\boldsymbol{\theta}),
		\label{eq:terminal_inverse_standardization}
	\end{equation}
	where $c^p$ and $s^p$ are the fixed initial-design standardization constants
	defined in Eqs.~\eqref{eq:response_center}--\eqref{eq:response_scale}.

	The final parameter estimate is obtained by minimizing the mean absolute
	mismatch between the physical posterior-mean response and the prescribed
	target:
	\begin{equation}
		\widehat{\boldsymbol{\theta}}
		\approx
		\arg\min_{\boldsymbol{\theta}\in\Theta}
		\frac{1}{d_y}
		\sum_{p=1}^{d_y}
		\left|
		\mu^{p,\mathrm{phys}}_B(\boldsymbol{\theta})
		-
		y_\star^p
		\right|.
		\label{eq:terminal_recommendation}
	\end{equation}
	For the present problem $d_y=20$. Since $d_y$ is fixed, minimizing this
	physical-response mean absolute error is equivalent to minimizing the
	corresponding unnormalized $L^1$ mismatch.

	Importantly, Eq.~\eqref{eq:terminal_recommendation} is optimized over the
	continuous admissible parameter domain rather than being restricted to
	previously evaluated simulator points. The terminal search begins from
	$1024$ scrambled-Sobol candidate points. The ten candidates with the lowest
	finite posterior-mean physical response error are used to initialize
	projected Adam refinement, with learning rate $0.05$ and at most $64$
	iterations per restart. Each iterate is projected onto the admissible
	parameter domain, and the best finite non-worsening state is retained. If no
	local refinement improves on its starting point, the best finite Sobol
	candidate is returned.

	The terminal optimization uses only the final GP posterior and therefore
	requires no additional simulator evaluations. It is also unconstrained by
	the compatibility condition used during sequential acquisition: the purpose
	of the terminal stage is solely to identify the parameter vector whose
	predicted physical response most closely matches the target after the
	simulation budget has been exhausted.

	The benchmark calculations use this continuous terminal policy. The
	evaluated-point alternative considered in
	Appendix~\ref{app:terminal} is used only as a terminal-domain ablation and is
	not part of the MUTACO algorithm.

	\subsection{Complete MUTACO algorithm}
	\label{app:mutaco_algorithm}

	The complete MUTACO procedure is summarized in
	Fig.~\ref{fig:mutaco_algorithm}. The algorithm combines the multitask-GP
	surrogate of Sec.~\ref{app:gp_surrogate}, the compatibility-guided
	acquisition of Sec.~\ref{app:gp_acquisition}, and the continuous terminal
	recommendation described above. The box is intended as a compact procedural
	summary; the preceding subsections define the corresponding quantities and
	objectives.

	\begin{figure*}[t]
		\centering
		\fbox{%
			\begin{minipage}{0.97\textwidth}
				\small

				\textbf{Algorithm: MUTACO target-matching search}

				\textbf{Input:}
				parameter domain $\Theta$, target response $\mathbf y^\star$,
				initial design size $n_0=16$, total simulator budget $B$,
				$\beta={\chi_{d_y}(0.95)}$,
				$\alpha=0.1$, and
				$\varepsilon_\Delta=10^{-8}$.

				\medskip

				\begin{minipage}[t]{0.48\linewidth}

					\textbf{1. Initial design.}
					Generate $n_0$ scrambled-Sobol points
					$\{\boldsymbol{\theta}_i\}_{i=1}^{n_0}\subset\Theta$
					and evaluate
					\[
					\mathbf y_i=\mathbf y(\boldsymbol{\theta}_i).
					\]
					Define
					\[
					\mathcal D_{n_0}
					=
					\{(\boldsymbol{\theta}_i,\mathbf y_i)\}_{i=1}^{n_0},
					\]
					and compute $c^p$ and $s^p$ from the initial design.
					These constants remain fixed thereafter

					\medskip

					\textbf{2. Fit the GP surrogate.}
					Fit the correlated multitask GP of
					Sec.~\ref{app:gp_surrogate} to the standardized responses.

					\medskip

					\textbf{3. Sequential acquisition.}
					For each remaining simulator evaluation:

					\smallskip

					\textbf{(a)} Generate a 64-point scrambled-Sobol screening set
					$\mathcal C_t\subset\Theta$. For each candidate compute
					\[
					\mathbf r_t(\boldsymbol{\theta})
					=
					\widetilde{\boldsymbol{\mu}}_t(\boldsymbol{\theta})
					-\widetilde{\mathbf y}^{\star},
					\]
					\[
					\lambda_t(\boldsymbol{\theta})
					=
					\lambda_{\max}
					\!\left[
					\widetilde{\boldsymbol{\Sigma}}_t(\boldsymbol{\theta})
					\right],
					\]
					and
					\[
					\Delta_t(\boldsymbol{\theta})
					=
					1-
					\frac{
						\|\mathbf r_t(\boldsymbol{\theta})\|_2^2
					}{
						\beta^2\lambda_t(\boldsymbol{\theta})
					}.
					\]

					The compatibility condition is
					\[
					\|\mathbf r_t(\boldsymbol{\theta})\|_2^2
					\le
					\beta^2\lambda_t(\boldsymbol{\theta}).
					\]

					\textbf{(b)} If at least one finite candidate satisfies
					$\Delta_t>\varepsilon_\Delta$, rank candidates using
					\[
					J_t(\boldsymbol{\theta})
					=
					\|\mathbf r_t(\boldsymbol{\theta})\|_2
					-\alpha\log\Delta_t(\boldsymbol{\theta}).
					\]
					Refine the four best screening points by bounded projected
					gradient descent with feasibility backtracking and select the
					best finite non-worsening result.

				\end{minipage}
				\hfill
				\begin{minipage}[t]{0.48\linewidth}

					\textbf{(c)} Otherwise, rank screening candidates by
					\[
					U_t(\boldsymbol{\theta})
					=
					\lambda_t(\boldsymbol{\theta}),
					\]
					and refine the four highest-ranked candidates by bounded projected
					Adam to approximately maximize
					\[
					\boldsymbol{\theta}_{t+1}
					\approx
					\arg\max_{\boldsymbol{\theta}\in\Theta}
					U_t(\boldsymbol{\theta}).
					\]

					\textbf{(d)} If the selected point coincides exactly with a
					previously evaluated point, return the highest-ranked finite
					unevaluated point from the screening set.

					\smallskip

					\textbf{(e)} Evaluate
					\[
					\mathbf y(\boldsymbol{\theta}_{t+1}),
					\]
					augment the dataset, and refit the multitask GP.

					\medskip

					\textbf{4. Final recommendation.}
					After exhausting the simulator budget, fit the final GP and return
					its posterior mean to physical response coordinates,
					\[
					\mu_B^{p,\mathrm{phys}}(\boldsymbol{\theta})
					=
					c^p+s^p\widetilde{\mu}_B^p(\boldsymbol{\theta}).
					\]
					The recommendation is
					\[
					\widehat{\boldsymbol{\theta}}
					\approx
					\arg\min_{\boldsymbol{\theta}\in\Theta}
					\frac{1}{d_y}
					\sum_{p=1}^{d_y}
					\left|
					\mu_B^{p,\mathrm{phys}}(\boldsymbol{\theta})
					-y_\star^p
					\right|.
					\]
					The continuous terminal optimization requires no additional
					simulator evaluations.

					\medskip

					\textbf{Output:}
					recommended parameter vector
					$\widehat{\boldsymbol{\theta}}$.

				\end{minipage}
			\end{minipage}%
		}
		\caption{
			Complete MUTACO target-matching procedure used in the benchmark calculations. The multitask GP surrogate, compatibility criterion,
			sequential acquisition rules, and terminal recommendation are defined
			in the preceding subsections. Additional numerical and reproducibility details are given in
			Appendix~\ref{app:numerics}.
		}
		\label{fig:mutaco_algorithm}
	\end{figure*}

	\section{Baseline Methods and Numerical Implementation}
	\label{app:baselines}

    \subsection{Scalar qLogNEI baseline}
\label{app:qlognei}

As an adaptive Bayesian-optimization baseline, we use logarithmic noisy
expected improvement (qLogNEI)~\cite{ament_logei}. A principal distinction
from MUTACO arises at the surrogate-modeling stage: qLogNEI reduces each
multicomponent simulator response to a scalar target-matching error before
fitting its surrogate, whereas MUTACO models the response components jointly
and retains their learned correlations and posterior uncertainty for
compatibility-aware acquisition.

For a simulator evaluation at parameter vector
$\boldsymbol{\theta}_i$, we define the physical response mean absolute error
\begin{equation}
    E_i
    =
    \frac{1}{d_y}
    \sum_{p=1}^{d_y}
    \left|
    y_i^p-y_\star^p
    \right|,
    \qquad d_y=20.
    \label{eq:qlognei_error}
\end{equation}
The scalar GP is trained on
\begin{equation}
    z_i=-E_i,
    \label{eq:qlognei_objective}
\end{equation}
so that maximizing the modeled objective is equivalent to minimizing the
physical response error. Because the forward simulator is stochastic,
$E_i$ is a noisy observation: repeated evaluations at the same
$\boldsymbol{\theta}$ need not give identical values.

The four parameter coordinates are normalized to the unit hypercube. The
scalar observations $z_i$ are standardized using the data available at each
GP refit. The surrogate is a single-output exact GP with a constant mean, an
RBF kernel with one length scale per parameter coordinate, and a learned
homoskedastic Gaussian observation-noise variance. Its hyperparameters are
obtained by maximizing the exact marginal likelihood.

The distinction between ordinary expected improvement and noisy expected
improvement is important in the present stochastic setting. Let
\begin{equation}
    \mathcal B_t
    =
    \{
    \boldsymbol{\theta}_1,\ldots,\boldsymbol{\theta}_t
    \}
\end{equation}
denote the previously evaluated baseline points. In ordinary expected
improvement with a fixed observed incumbent, one may define
\begin{equation}
    z_{\mathrm{best}}
    =
    \max_{i\leq t} z_i
\end{equation}
and evaluate improvement at a candidate $\boldsymbol{\theta}$ relative to
this fixed value,
\begin{equation}
    I_t(\boldsymbol{\theta})
    =
    \left[
    z(\boldsymbol{\theta})-z_{\mathrm{best}}
    \right]_+,
    \qquad
    [x]_+=\max(x,0).
\end{equation}

For noisy expected improvement, the latent objective values at the baseline
points are themselves uncertain. For each joint posterior sample $s$, let
\begin{equation}
    z_i^{(s)}
    =
    z^{(s)}(\boldsymbol{\theta}_i),
    \qquad i=1,\ldots,t,
\end{equation}
and let
\begin{equation}
    z_{\mathrm{cand}}^{(s)}
    =
    z^{(s)}(\boldsymbol{\theta})
\end{equation}
denote the corresponding sampled latent objective at the candidate. The
sample-dependent latent incumbent is then
\begin{equation}
    z_{\mathrm{best}}^{(s)}
    =
    \max_{i\leq t} z_i^{(s)},
\end{equation}
and the underlying Monte Carlo noisy-expected-improvement quantity for
$q=1$ is approximated by
\begin{equation}
    \mathrm{NEI}_t(\boldsymbol{\theta})
    \approx
    \frac{1}{N_{\mathrm{MC}}}
    \sum_{s=1}^{N_{\mathrm{MC}}}
    \left[
    z_{\mathrm{cand}}^{(s)}
    -
    z_{\mathrm{best}}^{(s)}
    \right]_+ .
    \label{eq:qnei_mc}
\end{equation}
Thus, unlike ordinary expected improvement with a fixed observed incumbent,
the identity of the effective best baseline point can change across posterior
samples as uncertainty in the previously evaluated points is marginalized.
This posterior re-ranking requires no additional simulator evaluations; it is
performed entirely within the fitted GP model. qLogNEI evaluates a
numerically stabilized logarithmic form of this noisy expected-improvement
acquisition.

We use $q=1$ and $N_{\mathrm{MC}}=256$ posterior Monte Carlo samples. All
previously evaluated parameter vectors enter the noisy baseline set, subject
to the baseline pruning used by the BoTorch implementation.

The acquisition function is optimized continuously over the full parameter
domain. Each sequential step uses 256 scrambled-Sobol raw candidates followed
by ten optimization restarts, with at most 64 optimizer iterations per
restart. The GP is refit after every new simulator evaluation. If acquisition
optimization produces a nonfinite, invalid, or previously evaluated point, the
same acquisition is evaluated on a 256-point scrambled-Sobol fallback set and
the highest-ranked finite unevaluated point is used.

After the simulation budget is exhausted, the final recommendation is obtained
from the posterior mean of the scalar GP. If
$\mu_B^{(-E)}(\boldsymbol{\theta})$ denotes the latent posterior mean expressed
on the original negative-MAE scale, then
\begin{equation}
    \widehat{\boldsymbol{\theta}}_{\mathrm{qLogNEI}}
    \approx
    \arg\max_{\boldsymbol{\theta}\in\Theta}
    \mu_B^{(-E)}(\boldsymbol{\theta}).
    \label{eq:qlognei_terminal}
\end{equation}
Thus, the terminal recommendation minimizes the GP-predicted physical
response error rather than simply returning the evaluated point with the
smallest noisy observed error. This terminal search is continuous over
$\Theta$, uses 1024 scrambled-Sobol raw points, ten optimization restarts,
and at most 64 iterations per restart, and requires no additional simulator
evaluations. The resulting recommendation is fixed before held-out
assessment; the subsequent held-out simulator evaluation is used only to
score that selected parameter vector and does not enter or revise the
recommendation.

Thus, qLogNEI and MUTACO are both adaptive, surrogate-based searches for the
same target response, but they organize the available information
differently. qLogNEI models a noisy scalar response-error objective and uses
posterior uncertainty through noisy expected improvement. MUTACO instead
models the correlated multicomponent response and uses its posterior
covariance to define target compatibility, barrier-based refinement, and an
uncertainty-driven exploration fallback. The benchmark therefore compares
the complete scalar qLogNEI strategy with the complete compatibility-aware
MUTACO strategy rather than isolating any single MUTACO design component.

	\subsection{Sobol search}
	\label{app:sobol}

	As a nonadaptive baseline, we use a scrambled Sobol sequence
	\cite{owen_scrambled_sobol} over the physical parameter domain defined in
	Eq.~\eqref{eq:inverse_bounds}. For each target and simulation budget, a
	length-$B$ Sobol sequence is generated in the unit four-cube $[0,1]^4$ and
	mapped affinely to the physical domain. If $u_i^a\in[0,1]$ denotes the
	$a$th coordinate of the $i$th Sobol point, then
	\begin{equation}
		\theta_i^a
		=
		\ell^a
		+
		\left(h^a-\ell^a\right)u_i^a,
		\qquad
		a=1,\ldots,4,
		\label{eq:sobol_mapping}
	\end{equation}
	where $\ell^a$ and $h^a$ are the lower and upper bounds of parameter
	coordinate $a$.

	The first $n_0=16$ Sobol points provide the common initial-design geometry
	used by MUTACO, scalar qLogNEI, and the Sobol baseline. For Sobol search, the
	remaining evaluations simply continue along the same predetermined sequence
	through point $B$. Thus, the full search contains 40 points for the $B=40$
	campaign and 100 points for the $B=100$ campaign.

	Unlike the adaptive methods, the Sobol trajectory does not depend on the
	observed simulator responses: no surrogate is fitted and no acquisition
	function is optimized. The simulator is nevertheless evaluated independently
	at every point, with a fresh stochastic context for each search evaluation.

	After all $B$ simulator evaluations have been completed, the Sobol
	recommendation is restricted to the evaluated parameter vectors. Let
	\begin{equation}
		E_i
		=
		\frac{1}{d_y}
		\sum_{p=1}^{d_y}
		\left|
		y_i^p-y_\star^p
		\right|
	\end{equation}
	denote the observed physical response MAE at the $i$th Sobol point. The final
	recommendation is
	\begin{equation}
		i^\star
		=
		\min\arg\min_{1\leq i\leq B} E_i,
		\qquad
		\widehat{\boldsymbol{\theta}}_{\mathrm{Sobol}}
		=
		\boldsymbol{\theta}_{i^\star}.
		\label{eq:sobol_terminal}
	\end{equation}
	The earliest evaluation is selected in the event of an exact tie.

	The Sobol terminal rule therefore uses only already evaluated simulator
	points and requires no surrogate fit, continuous optimization, or additional
	simulator calls.

	\subsection{Common initialization and terminal policies}
	\label{app:common_policies}

	All three primary methods use the same initial-design geometry within a given
	target and simulation budget. For each target-budget pair, a length-$B$
	scrambled-Sobol design is generated over the four-dimensional parameter
	domain, and its first
	\begin{equation}
		n_0=16
	\end{equation}
	points are supplied to MUTACO, scalar qLogNEI, and Sobol. Thus, the methods
	begin from identical parameter locations.

	The corresponding simulator observations are not shared. Each method
	evaluates the common initial parameter points independently using its own
	method-addressed stochastic simulator seeds. Consequently, the initial
	geometry is common across methods, whereas the realized disorder,
	initial-state, and stochastic-noise samples are method specific. This distinction
	is maintained throughout the adaptive search.

	The total simulator budgets are
	\begin{equation}
		B=40
		\qquad\text{and}\qquad
		B=100.
	\end{equation}
	Accordingly, the adaptive methods use
	\begin{equation}
		16+24=40
	\end{equation}
	and
	\begin{equation}
		16+84=100
	\end{equation}
	simulator evaluations, respectively. For the Sobol baseline, the same
	accounting corresponds to the first 16 common-design points followed by 24
	or 84 additional points from the predetermined Sobol sequence.

	The $B=40$ and $B=100$ studies are independent campaigns rather than
	checkpoints along a common trajectory. They use separate design and
	randomness roots, and their initial Sobol designs therefore need not coincide
	numerically.

	Although the three methods use the same search budgets, their terminal
	recommendation policies are method specific. These policies are summarized
	in Table~\ref{tab:terminal_policies}. In all cases, terminal recommendation
	occurs only after the search budget has been exhausted and consumes no
	additional simulator evaluations.

	\begin{table*}[t]
		\caption{
			Terminal recommendation policies used in the 120-target benchmark
			comparison. All terminal procedures are applied after the corresponding
			search budget is exhausted and consume no additional simulator calls.
			The methods address the same response-matching task but use terminal
			criteria appropriate to the information represented by their respective
			search procedures.
		}
		\label{tab:terminal_policies}
		\begin{ruledtabular}
			\begin{tabular}{lccc}
				\textbf{Method}
				&
				\textbf{Terminal recommendation criterion}
				&
				\textbf{Search domain}
				&
				\textbf{Numerical policy}
				\\
				\hline
				MUTACO
				&
				MAE of posterior-mean physical response
				&
				Continuous $\Theta$
				&
				\shortstack{1024 Sobol starts,\\10 projected-Adam restarts}
				\\
				Scalar qLogNEI
				&
				Posterior mean of scalar objective $-E$
				&
				Continuous $\Theta$
				&
				1024 Sobol starts, 10 L-BFGS-B restarts
				\\
				Sobol
				&
				Observed physical response MAE $E$
				&
				Evaluated points only
				&
				Best observed point
			\end{tabular}
		\end{ruledtabular}
	\end{table*}

	For MUTACO, the final multitask-GP posterior mean is transformed back to
	physical response coordinates, and the mean absolute mismatch between that
	posterior-mean response and the prescribed target is minimized continuously,
	as described in Sec.~\ref{app:gp_terminal}. For scalar qLogNEI, each simulator
	evaluation is instead reduced to the scalar response-matching objective
	$-E$ before surrogate modeling. Its terminal recommendation is therefore
	obtained by continuously maximizing the final scalar-GP posterior mean,
	Eq.~\eqref{eq:qlognei_terminal}. Both surrogate-based methods search the same
	admissible parameter domain and use 1024 scrambled-Sobol raw points, ten
	optimization restarts, and at most 64 optimizer iterations. Their terminal
	criteria are nevertheless not mathematically identical because MUTACO models
	the multicomponent response itself, whereas qLogNEI models its scalar mismatch
	to the target.

	The Sobol baseline constructs no surrogate and therefore has no corresponding
	continuous posterior objective to optimize. Its recommendation is restricted
	to parameter vectors actually evaluated during the search and is the point
	with the smallest observed physical response MAE, as defined in
	Eq.~\eqref{eq:sobol_terminal}. The terminal policy is thus part of each
	complete search-and-recommendation procedure and is chosen consistently with
	the information represented by that procedure rather than being artificially
	forced to be identical across methods. The benchmark holds fixed the
	simulator-call budget and evaluates the resulting recommendations using the
	same held-out physical response metric; it should therefore be interpreted as
	a comparison of the complete methods rather than of their acquisition rules
	in isolation. The contribution of MUTACO's continuous terminal search is
	examined separately by the controlled terminal-domain ablation in
	Appendix~\ref{app:terminal}.

	\begin{table*}[t]
		\caption{
			Numerical design and simulator-call accounting for the 120-target
			benchmark. The $B=40$ and $B=100$ calculations are independent
			fixed-budget experiments.
		}
		\label{tab:benchmark_protocol}
		\begin{ruledtabular}
			\begin{tabular}{ll}
				\textbf{Quantity}
				&
				\textbf{Benchmark specification}
				\\
				\hline
				Target instances
				&
				120 independently generated target responses
				\\
				Parameter dimension
				&
				$d_\theta=4$
				\\
				Response dimension
				&
				$d_y=20$
				\\
				Parameter domain
				&
				$\Theta=[0.5,2]^3\times[0,1]$
				\\
				Response-matching methods
				&
				MUTACO, scalar qLogNEI, scrambled Sobol
				\\
				Simulator budgets
				&
				$B=40,\ 100$ calls per method and target
				\\
				Initial design
				&
				$n_0=16$ parameter locations
				\\
				Sequential evaluations
				&
				24 for $B=40$; 84 for $B=100$
				\\
				Initial parameter geometry
				&
				Common across the three response-matching methods
				\\
				Search randomness
				&
				Independent simulator realizations across methods and evaluations
				\\
				Target construction
				&
				Performed before inverse search; excluded from $B$
				\\
				Terminal recommendation
				&
				Performed after the search; no additional simulator calls
				\\
				Held-out assessment
				&
				One simulator evaluation per method, target, and budget
				\\
				Held-out randomization
				&
				Method-independent stochastic context paired across methods
				\\
				Assessment accounting
				&
				Excluded from the search budget $B$
				\\
				Budget relation
				&
				$B=40$ and $B=100$ treated as independent experiments
			\end{tabular}
		\end{ruledtabular}
	\end{table*}

	\begin{figure*}[t]
		\centering
		\includegraphics[width=0.82\textwidth]
		{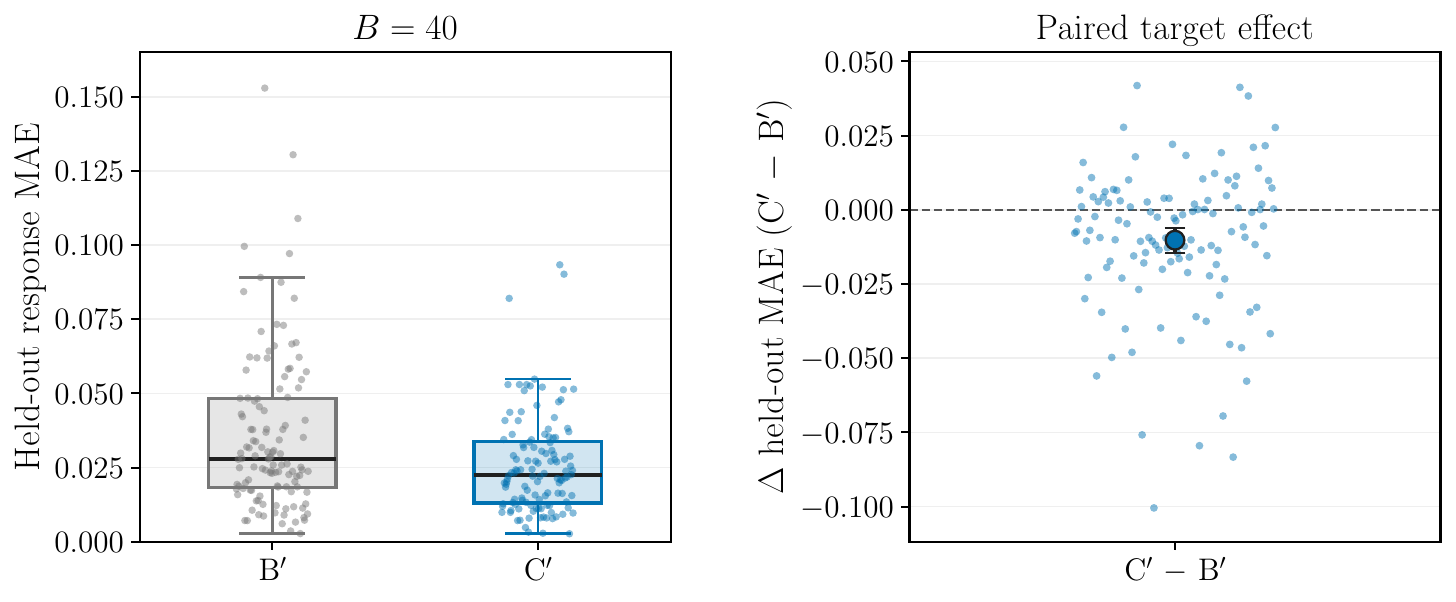}
		\caption{
			Terminal-domain ablation for the $B=40$ MUTACO population.
			B$^\prime$ restricts the final recommendation to parameter vectors
			already evaluated during the sequential search, whereas C$^\prime$
			optimizes the same final GP posterior-mean physical response MAE
			continuously over the full admissible parameter domain.
			All other aspects of the completed MUTACO trajectories and final
			surrogates are held fixed.
		}
		\label{fig:terminal_domain_ablation}
	\end{figure*}

	\subsection{{SBI implementation and point recommendations}}
	
	The NPE, SNPE, NLE, and NRE baselines are implemented using the
	\texttt{sbi} toolkit~\cite{BoeltsDeistler_sbi_2025,tejero-cantero2020sbi}.
	
	{All four SBI methods use a uniform prior on
	$\Theta=[0.5,2]^3\times[0,1]$. NPE uses one prior-predictive round of $B$
	simulator calls; SNPE-C/APT uses three rounds with allocations
	$(14,13,13)$ for $B=40$ and $(34,33,33)$ for $B=100$.
	Within each target--budget pair, NLE and NRE train on the same $B$-call
	prior-predictive dataset. NPE, SNPE, and NLE use masked autoregressive
	flows; NRE uses a two-hidden-layer multilayer perceptron. Training uses a
	learning rate of $5\times10^{-4}$, a $20\%$ validation fraction, batches
	of at most 64, early-stopping patience of 20 epochs, and at most 200 epochs.}

	{For each SBI method, the final point recommendation is selected from
	$10^4$ posterior samples as the in-domain sample with the largest finite
	posterior score in physical parameter coordinates; exact ties favor the
	earliest sample. This selection uses no additional simulator calls.
	One subsequent held-out simulator evaluation assesses the fixed
	recommendation and does not revise it.}

	\subsection{Numerical implementation and reproducibility}
	\label{app:numerics}

	All benchmark calculations use fixed, reproducible random seeds determined by
	the scientific configuration and experiment coordinates, with independent
	random-number streams for design generation, inference, terminal optimization,
	and held-out assessment. The 120-target cohort was generated with master seed
	$92020120$ and fixed before any inverse-search calculations; the benchmark
	records retain the generating parameters, target responses, and information
	needed to reproduce the randomization.

	During the search itself, every simulator evaluation generates a fresh
	stochastic realization. In particular, the quenched disorder, oscillator
	initial state, and stochastic noise are regenerated for each evaluation. Common
	random numbers are therefore not used between successive adaptive queries or
	between competing methods. Even at the common initial parameter locations,
	MUTACO, scalar qLogNEI, and Sobol use separately generated simulator
	realizations. The only quantity shared at initialization is the common set of
	initial parameter locations described in Sec.~\ref{sec:randomization}.

	Random-number generation is isolated by numerical operation. Scrambled-Sobol
	designs use dedicated Sobol engines, while the stochastic simulator uses
	separately seeded generators for quenched disorder, initial conditions, and
	stochastic noise. GP fitting, acquisition optimization, and terminal optimization
	likewise isolate and restore the ambient Torch random-number state. As a
	result, random choices made by one numerical operation do not alter the random
	stream assigned to another.

	The held-out assessment uses a different randomization policy. All search
	trajectories and terminal recommendations are completed before held-out
	evaluation begins. For each target and budget, a single method-independent
	stochastic context is then generated and reused to assess all three
	recommendations. The methods therefore receive separate simulator calls but
	identical held-out disorder, initial-state, and stochastic-noise realizations.
	This yields a paired comparison across methods without introducing common
	random numbers into the adaptive searches themselves.

	Held-out performance is evaluated using the physical response error defined
	in Eq.~\eqref{eq:heldout_response_error}. One held-out simulator evaluation
	is performed for each method, target, and budget, with the stochastic context
	paired across methods. These evaluations occur only after the recommendations
	have been fixed and are not counted against the search budget.

	Because the benchmark targets are generated from known ensemble parameters,
	we additionally report the parameter-space diagnostic defined in
	Eq.~\eqref{eq:parameter_error}, evaluated directly in the four unnormalized
	physical parameter coordinates. The generating vector
	$\boldsymbol{\theta}^{\star}$ is used only for evaluation and is not available
	to any of the search methods.

	For a single target, the three response-matching methods therefore require
	$3B$ search evaluations in total: $B$ each for MUTACO, scalar qLogNEI, and
	Sobol. The three paired held-out evaluations are additional assessment calls
	and do not enter this budget. Terminal optimization for MUTACO and qLogNEI
	uses only the fitted surrogate and therefore requires no additional simulator
	evaluations, whereas the Sobol recommendation is selected directly from its
	evaluated points.

	The benchmark calculations used Python~3.11.15,
	PyTorch~2.12.1+cpu, BoTorch~0.18.1~\cite{balandat2020botorch}, and
	GPyTorch~1.15.2~\cite{gardner2018gpytorch}. Scientific tensors
	were evaluated on the CPU in single precision. GP hyperparameters were fit
	by exact marginal-likelihood optimization. The scalar qLogNEI implementation
	uses BoTorch's bounded fit-recovery procedure and acquisition-optimizer retry
	mechanism, together with the deterministic finite-candidate fallback
	described in Sec.~\ref{app:qlognei}. MUTACO uses the bounded fitting and
	proposal-recovery procedures described in Appendix~\ref{app:gp}.

	The $B=40$ and $B=100$ benchmark experiments were performed independently,
	with separate design, inference, terminal, and assessment seed roots. They
	should therefore be interpreted as independent fixed-budget experiments rather
	than as checkpoints along a common search trajectory.

	The complete benchmark protocol is summarized in
	Table~\ref{tab:benchmark_protocol}.

	\section{Terminal-Domain Ablation}
	\label{app:terminal}

	The primary MUTACO procedure separates sequential data acquisition from the
	terminal recommendation. After the simulation budget has been exhausted, the
	primary method searches continuously over the admissible parameter domain for
	the minimum of the final GP posterior-mean physical response MAE, as described
	in Sec.~\ref{app:gp_terminal}. We performed a controlled ablation to determine
	how much of the final response accuracy is attributable to this continuous
	terminal search rather than to the preceding acquisition trajectory.

	For each target, we refit the final GP to the 40 observations collected during the completed primary MUTACO search. The evaluated-point and continuous terminal policies then use the same refitted surrogate, target, and objective, differing only in whether the terminal candidate is restricted to previously evaluated points or optimized continuously over  $\Theta$. We compare two
	terminal policies:

	\begin{align}
		\text{B}^{\prime}:&\quad
		\widehat{\boldsymbol{\theta}}_{\mathrm{eval}}
		=
		\arg\min_{\boldsymbol{\theta}\in\mathcal D_B}
		E_{\mathrm{GP}}(\boldsymbol{\theta}),
		\label{eq:terminal_Bprime}
		\\[4pt]
		\text{C}^{\prime}:&\quad
		\widehat{\boldsymbol{\theta}}_{\mathrm{cont}}
		\approx
		\arg\min_{\boldsymbol{\theta}\in\Theta}
		E_{\mathrm{GP}}(\boldsymbol{\theta}),
		\label{eq:terminal_Cprime}
	\end{align}
	where $\mathcal D_B$ denotes the set of parameter vectors evaluated during
	the search and
	\begin{equation}
		E_{\mathrm{GP}}(\boldsymbol{\theta})
		=
		\frac{1}{d_y}
		\sum_{p=1}^{d_y}
		\left|
		\mu_B^{p,\mathrm{phys}}(\boldsymbol{\theta})
		-
		y_\star^p
		\right|
	\end{equation}
	is the same final posterior-mean physical response MAE used by the primary
	terminal policy.

	Neither policy requires an additional simulator evaluation.

	Fig.~\ref{fig:terminal_domain_ablation} shows the resulting held-out
	response errors for the evaluated-point and continuous terminal policies
	across the 120 $B=40$ targets.

	Across the 120 targets, the evaluated-point policy B$^\prime$ gives a mean
	held-out physical response MAE of $3.60\times 10^{-2}$ and a median of $2.80\times 10^{-2}$,
	whereas continuous terminal optimization C$^\prime$ gives a mean of $2.57\times 10^{-2}$
	and a median of $2.24\times 10^{-2}$. The paired mean difference in this held-out MAE,
	\begin{equation}
		\overline{E_{\mathrm{C}^{\prime}}-E_{\mathrm{B}^{\prime}}}
		=
		-1.02\times 10^{-2},
	\end{equation}
	corresponds to a $28.5\%$ reduction in mean held-out physical response MAE. The
	paired 95\% interval for the difference is
	\begin{equation}
		[-1.47,\,-0.603]\times 10^{-2}.
	\end{equation}
	Continuous optimization gives the lower terminal error for 75 of the 120
	targets, while the evaluated-point policy is better for 45; no exact ties
	occur.

	This ablation shows that the continuous terminal search provides a measurable
	improvement beyond simply selecting among points visited during sequential
	acquisition. At the same time, it does not alter the acquisition trajectory,
	the simulation budget, or the fitted final GP. The result therefore isolates
	the effect of the terminal recommendation domain rather than constituting a
	comparison of acquisition strategies.

	\bibliography{bandits_v6}% Produces the bibliography via BibTeX.

\end{document}